\documentclass[12pt]{iopart}
\usepackage[dvips,final]{graphicx}
\usepackage{latexsym}

\usepackage{iopams}

\newcommand{\be}{\begin{equation}}
\newcommand{\ee}{\end{equation}}
\newcommand{\bea}{\begin{eqnarray}}
\newcommand{\eea}{\end{eqnarray}}

\def\be{\begin{equation}}
\def\ee{\end{equation}}
\def\ba{\begin{eqnarray}}
\def\ea{\end{eqnarray}}
\def\l{\left}
\def\r{\right}
\def\f{\frac}

\def\mn{\mu\nu}

\def\gapp{\mathrel{\raise.3ex\hbox{$>$}\mkern-14mu
              \lower0.6ex\hbox{$\sim$}}}

\def\lapp{\mathrel{\raise.3ex\hbox{$<$}\mkern-14mu
              \lower0.6ex\hbox{$\sim$}}}

\newcommand{\mpl}{M_p}

\def\f{\frac}
\begin{document}

\title[Approaches to Understanding Cosmic Acceleration]{Approaches to Understanding Cosmic Acceleration}

\author{Alessandra Silvestri$^1$ and Mark Trodden$^2$}

\address{$^1$Department of Physics and Kavli Institute for Astrophysics and Space Research, MIT, 77 Massachusetts Ave., Cambridge, MA 02139, USA.}

\address{$^2$Center for Particle Cosmology, Department of Physics and Astronomy, David Rittenhouse Laboratories, University of Pennsylvania, Philadelphia, PA 19104, USA.}

\ead{asilvest@mit.edu, trodden@physics.upenn.edu}

\begin{abstract}
Theoretical approaches to explaining the observed acceleration of the universe are reviewed. We briefly discuss the evidence for cosmic acceleration, and the implications for standard General Relativity coupled to conventional sources of energy-momentum. We then address three broad methods of addressing an accelerating universe: the introduction of a cosmological constant, its problems and origins; the possibility of dark energy, and the associated challenges for fundamental physics; and the option that an infrared modification of general relativity may be responsible for the large-scale behavior of the universe.
\end{abstract}


\maketitle

\section{Introduction}
\label{sec:introduction}
The development of General Relativity almost 90 years ago provided not only a new way to understand gravity, but heralded the dawn of an entirely new branch of science - cosmology. Of course, a discussion of the visible contents of the night sky was not a new development - these objects and phenomena have forever fascinated civilizations. However, a coherent theoretical framework within which to explore their distribution and evolution was lacking before Einstein's crucial insight - that space and time are themselves players in the cosmic drama. Hubble's 1921 observation that distant galaxies are receding from ours at speeds proportional to their distances, in agreement with Friedmann's corresponding solution to General Relativity, then provided the first experimental confirmation of this new science.

Almost a century has passed since the beginning of this era, and in the intervening years increasingly accurate predictions of this model of the cosmos, supplemented only by the presence of a dark matter component, have been confronted with, and spectacularly passed, a host of detailed tests - the existence of the Cosmic Microwave Background Radiation (CMB); the abundances of the light elements through Big Bang Nucleosynthesis (BBN); the formation of structure under gravitational instability; the small temperature anisotropies in the CMB; the structure of gravitational lensing maps; and many more. Many of these tests are highly nontrivial and provide remarkable support for the overall big bang model.

All this held true until approximately a decade ago, when the first evidence for cosmic acceleration was reported by two groups using the lightcurves of type Ia supernovae to construct an accurate Hubble diagram out to redshift greater than unity. As we will describe in some detail, the natural expectation is that an expanding universe, evolving under the rules of GR, and populated by standard matter sources, will undergo deceleration - the expansion rate should slow down as cosmic time unfolds. Amazingly, that initial data, supplemented over the last decade by dozens of further, independent observations, showed that the universe is speeding up - what we observe is not deceleration, but cosmic acceleration!

The goal of this article is to provide an overview of what is meant by cosmic acceleration, some proposed theoretical approaches to understanding the phenomenon, and how upcoming observational data may cast crucial light on these. We have not attempted to be completely comprehensive, but have tried to provide a flavor of the current research landscape, focusing on the better known approaches, and highlighting the challenges inherent in constructing a fundamental physics understanding of the accelerating universe. There are a number of excellent recent review articles on dark energy, which the reader might find as a useful complement to the present paper~\cite{Copeland:2006wr,Linder:2008pp,Frieman:2008sn,Caldwell:2009ix}.
 
The article is organized as follows. In the next section we provide a brief introduction to the relevant cosmological solutions to General Relativity, with perfect fluid sources. In section~\ref{sec:observations} we review the observational evidence, primarily from observations made during the last decade or so, for an accelerating universe.  In section~\ref{sec:cosmoconstant} we then discuss the minimal possibility that Einstein's cosmological constant is responsible for late-time cosmic acceleration, and discuss the challenges the required magnitude of this parameter raises for particle physics. In~\ref{sec:darkenergy} we then explore a class of dynamical {\it dark energy} models, in which the universe accelerates due to the evolution of a new component of the cosmic energy budget, under the assumption that the cosmological constant is negligibly small. In a similar vein, in~\ref{sec:MGR} we then consider the possibility that the universe contains only standard model and dark matter sources of energy and momentum, but accelerates because GR is modified at large distances, leading to new self-accelerating solutions. Finally, before concluding, in~\ref{sec:tests} we then consider the challenge of distinguishing among these possible explanations for cosmic acceleration, and discuss how upcoming observational missions may help us to better understand this phenomena.

A note on conventions. Throughout this article we use units in which $\hbar=c=1$, adopt the $(-,+,+,+)$ signature and define the reduced Planck mass by $M_p^{-2}\equiv \kappa^{2}= 8\pi G$.


\section{Essential Elements of Background Cosmology}
\label{sec:background}

Our goal here is to lay out the bare minimum for understanding the background evolution of the universe. By {\it background} in this context, we mean the dynamics pertaining to the homogeneous and isotropic description of spacetime, valid on the very largest scales, without reference to spatial perturbations of either the metric or matter fields\footnote{We follow closely the brief discussion in~\cite{Trodden:2004st}}.

The most general homogeneous and isotropic metric ansatz is the Friedmann, Robertson-Walker (FRW) form
\begin{equation}
\label{FRWmetric}
ds^2=-dt^2 +a^2(t)\left[\frac{dr^2}{1-kr^2}+r^2\left(d\theta^2+\sin^2\theta d\phi^2\right)\right] \ ,
\end{equation}
where $k$ describes the curvature of the spatial sections (slices at constant cosmic time) and $a(t)$ is referred to as the {\it scale factor} or the universe. 
Without loss of generality, we may normalize $k$ so that $k=+1$ corresponds to positively curved spatial sections (locally isometric to 3-spheres); $k=0$ corresponds to local flatness, and $k=-1$ corresponds to negatively curved (locally hyperbolic) spatial sections. These local definitions say nothing about the global topology of the spatial sections, which may be that of the covering spaces -- a 3-sphere, an infinite plane or a 3-hyperboloid -- but it need not be, as topological identifications under freely-acting subgroups of the isometry group of each manifold are allowed. As a specific example, the $k=0$ spatial geometry could apply just as well to a 3-torus as to an infinite plane.

Note that we have not chosen a normalization such that $a_0=1$.
We are not free to do this and to simultaneously normalize $|k|=1$,
without including explicit factors of the current scale factor in
the metric.  In the flat case, where $k=0$, we can safely choose
$a_0=1$.

At cosmic time $t$, the physical distance from the origin to an object at radial coordinate r is given by $d(t)=a(t)r$. The recessional velocity of such an object due to the expansion of the universe is then given by
\begin{equation}
v(t)=H(t)d(t) \ ,
\end{equation}
where $H(t)\equiv {\dot a}/a$ (an overdot denotes a derivative with respect to $t$) is the Hubble parameter, the present day value of which we refer to as the {\it Hubble constant} $H_0$. 

The Hubble parameter is a useful function through which to parametrize the expansion rate of the universe. As we shall encounter later, it is convenient to define a second function, describing the rate at which the expansion rate is slowing down or speeding up - the {\it deceleration parameter}, defined as
\be
q(t) \equiv -\frac{{\ddot a}a}{{\dot a}^2} \ .
\label{decelerationparameter}
\ee

\subsection{Dynamics:  The Friedmann Equations}

The FRW metric is merely an ansatz, arrived at by requiring homogeneity and isotropy of spatial sections.  The unknown
function $a(t)$ is obtained by solving the differential equations obtained by substituting the FRW ansatz into the Einstein equation 
\be
R_{\mu\nu}-{1\over 2}Rg_{\mu\nu} =8\pi G T_{\mu\nu} \ .
  \label{einstein}
\ee

The energy-momentum tensors $T_{\mu\nu}$ describes the matter content of the universe.  
It is often appropriate to adopt the perfect fluid form for this
\begin{equation}
\label{perfectfluid}
T_{\mu\nu} = (\rho + p)U_\mu U_\nu + p g_{\mu\nu}\ ,
\end{equation}
where $U^{\mu}$ is the fluid four-velocity, $\rho$ is the energy
density in the rest frame of the fluid and $p$ is the pressure in that
frame.  

Substituting~(\ref{FRWmetric}) and~(\ref{perfectfluid}) into~(\ref{einstein}), one
obtains the Friedmann equation
\begin{equation}
\label{Friedmann}
H^2 \equiv \left(\frac{{\dot a}}{a}\right)^2=\frac{8\pi G}{3}\sum_i \rho_i -\frac{k}{a^2} \ ,
\end{equation}
and 
\begin{equation}
\label{2ndeinsteineqn}
\frac{\ddot a}{a} +\frac{1}{2}\left(\frac{{\dot a}}{a}\right)^2=-4\pi G\sum_i p_i -\frac{k}{2a^2} \ .
\end{equation}
Here $i$ indexes all different possible types of energy in the
universe. The Friedmann equation is a constraint equation, in the sense that we
are not allowed to freely specify the time derivative $\dot{a}$; it is
determined in terms of the energy density and curvature. The second
equation an evolution equation.

We may combine~(\ref{Friedmann}) and~(\ref{2ndeinsteineqn}) to obtain the {\it acceleration equation}
\begin{equation}
\label{acceleration}
\frac{{\ddot a}}{a}=-\frac{4\pi G}{3}\sum_i \left(\rho_i +3p_i \right) \ ,
\end{equation}
which will prove central to the subject of this review.

In fact, if we know the magnitudes and evolutions of the different
energy density components $\rho_i$, the Friedmann equation
(\ref{Friedmann}) is sufficient to solve for the evolution uniquely.
The acceleration equation is conceptually useful, but rarely invoked
in calculations.

The Friedmann equation relates the rate of increase of the scale
factor, as encoded by the Hubble parameter, to the total energy
density of all matter in the universe. We may use the Friedmann
equation to define, at any given time, a critical energy density,
\begin{equation}
\label{criticaldensity}
\rho_c\equiv \frac{3H^2}{8\pi G} \ ,
\end{equation}
for which the spatial sections must be precisely flat ($k=0$). 
We then define the density parameter
\begin{equation}
\label{omega}
\Omega_{\rm total} \equiv \frac{\rho}{\rho_c} \ ,
\end{equation}
which allows us to relate the total energy density in the universe to
its local geometry via
\begin{eqnarray}
\Omega_{\rm total}>1 & \Leftrightarrow & k=+1 \nonumber \\
\Omega_{\rm total}=1 & \Leftrightarrow & k=0 \\
\Omega_{\rm total}<1 & \Leftrightarrow & k=-1 \nonumber \ .
\end{eqnarray}
It is often convenient to define the fractions of the critical energy
density in each different component by
\begin{equation}
\Omega_i=\frac{\rho_i}{\rho_c} \ .
\end{equation}

Energy conservation is expressed in GR by the vanishing of the
covariant divergence of the energy-momentum tensor,
\be
  \nabla_\mu T^{\mu\nu} = 0\ .
\ee
For the FRW metric (\ref{FRWmetric}) and a perfect-fluid energy-momentum tensor (\ref{perfectfluid}) this yields
a single energy-conservation equation,
\begin{equation}
\label{energyconservation}
{\dot \rho} + 3H(\rho+p)=0 \ .
\end{equation}
This equation is actually not independent of the Friedmann and
acceleration equations, but is required for consistency.  It implies
that the expansion of the universe (as specified by $H$) can lead to
local changes in the energy density.  Note that there is no notion of
conservation of ``total energy,'' as energy can be interchanged
between matter and the spacetime geometry.

One final piece of information is required before we can think about
solving our cosmological equations: how the pressure and energy
density are related to each other.  Within the fluid approximation
used here, we may assume that the pressure is a single-valued function
of the energy density $p=p(\rho)$. It is often convenient to define an
equation of state parameter, $w$, by
\be
p= w\rho\ .
\ee 
This should be thought of as the instantaneous definition of the 
parameter $w$; it does not represent the full equation of state, which would
be required to calculate the behavior of fluctuations.
Nevertheless, many useful cosmological matter sources do
obey this relation with a constant value of $w$. For example, $w=0$ 
corresponds to pressureless matter, or dust -- any collection of
massive non-relativistic particles would qualify.  Similarly,
$w=1/3$ corresponds to a gas of radiation, whether it be actual
photons or other highly relativistic species.

A constant $w$ leads to a great simplification in solving our
equations. In particular, using~(\ref{energyconservation}), we see
that the energy density evolves with the scale factor according to
\begin{equation}
\label{energydensity}
\rho(a) \propto \frac{1}{a(t)^{3(1+w)}} \ .
\end{equation} 
Note that the behaviors of dust ($w=0$) and radiation ($w=1/3$) are
consistent with what we would have obtained by more heuristic
reasoning. Consider a fixed {\it comoving} volume of the universe -
i.e. a volume specified by fixed values of the coordinates, from which
one may obtain the physical volume at a given time $t$ by multiplying
by $a(t)^3$. Given a fixed number of dust particles (of mass $m$)
within this comoving volume, the energy density will then scale just
as the physical volume, i.e. as $a(t)^{-3}$, in agreement
with~(\ref{energydensity}), with $w=0$.

To make a similar argument for radiation, first note that the
expansion of the universe (the increase of $a(t)$ with time) results
in a shift to longer wavelength $\lambda$, or a {\it redshift}, of
photons propagating in this background. A photon emitted with
wavelength $\lambda_e$ at a time $t_e$, at which the scale factor is
$a_e\equiv a(t_e)$ is observed today ($t=t_0$, with scale factor
$a_0\equiv a(t_0)$) at wavelength $\lambda_o$, obeying
\begin{equation}
\label{redshift}
\frac{\lambda_o}{\lambda_e}=\frac{a_0}{a_e}\equiv 1+z \ .
\end{equation}
The redshift $z$ is often used in place of the scale factor.  Because
of the redshift, the energy density in a fixed number of photons in a
fixed comoving volume drops with the physical volume (as for dust) and
by an extra factor of the scale factor as the expansion of the
universe stretches the wavelengths of light. Thus, the energy density
of radiation will scale as $a(t)^{-4}$, once again in agreement
with~(\ref{energydensity}), with $w=1/3$.

Thus far, we have not included a cosmological constant $\Lambda$ in
the gravitational equations. This is because it is equivalent to treat
any cosmological constant as a component of the energy density in the
universe. In fact, adding a cosmological constant $\Lambda$ to
Einstein's equation is equivalent to including an energy-momentum
tensor of the form
\be
  T_{\mu\nu} = -{\Lambda \over 8\pi G} g_{\mu\nu}\ .
\ee
This is simply a perfect fluid with energy-momentum 
tensor~(\ref{perfectfluid}) with
\begin{eqnarray}
\rho_{\Lambda} & = & \frac{\Lambda}{8\pi G} \nonumber \\
p_{\Lambda} & = & -\rho_{\Lambda} \ , 
\end{eqnarray}
so that the equation-of-state parameter is
\be
  w_{\Lambda}=-1\ .
\ee 
This implies that the energy density is constant,
\be
 \rho_\Lambda = {\rm constant}\ .
\ee
Thus, this energy is constant throughout spacetime; we say that
the cosmological constant is equivalent to {\it vacuum
energy}. 

Similarly, it is sometimes useful to think of any nonzero spatial curvature as yet another component of the 
cosmological energy budget, obeying
\begin{eqnarray}
\rho_{\rm curv} & = & -\frac{3k}{8\pi Ga^2} \nonumber \\
p_{\rm curv} & = & \frac{k}{8\pi Ga^2}\ , 
\end{eqnarray}
so that 
\be
w_{\rm curv}=-1/3 \ .
\ee
It is not an energy density, of course; $\rho_{\rm curv}$ is
simply a convenient way to keep track of how much energy density
is {\rm lacking}, in comparison to a flat universe.

\subsection{Flat Universes}
Analytically, it is much easier to find exact solutions to cosmological equations of motion when $k=0$. We are able to make full use of this convenience since, as we shall touch on in the next section, the combined results of modern cosmological observations show the present day universe to be extremely spatially flat. This can be seen, for example, from precision observations of the cosmic microwave background radiation and independent measures of the Hubble expansion rate.

In the case of flat spatial sections and a constant equation of state parameter $w$, we may exactly solve the Friedmann equation 
(\ref{energydensity}) to obtain
\begin{equation}
\label{flatsolution}
a(t)=a_0 \left(t\over t_0\right)^{2/3(1+w)} \ ,
\end{equation}
where $a_0$ is the scale factor today, unless $w=-1$, 
in which case one obtains $a(t) \propto e^{Ht}$.
Applying this result to some of our favorite energy density sources yields table~\ref{sourcestable}.
\begin{table}[t]
\begin{center}
\begin{tabular}{l|l|l}
Type of Energy & $\rho(a)$ & $a(t)$ \\ \hline
Dust & $a^{-3}$ & $t^{2/3}$ \\
Radiation & $a^{-4}$ & $t^{1/2}$ \\
Cosmological Constant & constant & $e^{Ht}$ \\
\end{tabular}
\end{center}
\caption{A summary of the behaviors of the most important sources of energy density in cosmology.  The behavior of the scale factor applies
to the case of a flat universe; the behavior of the energy densities
is perfectly general.}
\label{sourcestable}
\end{table}

Note that the matter- and radiation-dominated flat universes begin
with $a=0$; this is a singularity, known as the Big Bang.  We can
easily calculate the age of such a universe:
\be
  t_0 = \int_0^1 {da \over aH(a)} =
  {2\over 3(1+w)H_0}\ .
\ee
Unless $w$ is close to $-1$, it is often useful to approximate this
answer by
\be
  t_0 \sim H_0^{-1}\ .
\ee
It is for this reason that the quantity $H_0^{-1}$ is known as the
{\it Hubble time}, and provides a useful estimate of the time scale
for which the universe has been around.


\section{Observational Evidence for Cosmic Acceleration}
\label{sec:observations}

An important breakthrough in cosmology occurred in the late '90s with the measurements of type Ia Supernovae (SNeIa) by the High-Z Supernova team~\cite{Riess:1998cb} and the Supernova Cosmology Project~\cite{Perlmutter:1998np}. As we review in this section, the data from these surveys, as well as from complementary probes, provide strong evidence that the universe has recently entered a phase of accelerated expansion.

\subsection{Type Ia Supernovae}
Type Ia supernovae are stellar explosions that occur as a white dwarf, onto which mass is gradually accreting from a companion star, approaches the Chandrasekhar limit~\cite{Chandrasekhar:1931ih} (of about $1.4$ solar masses) and the density and temperature in its core reach the ignition point for carbon and oxygen. This begins a nuclear flame that fuses much of the star up to iron. They are extremely bright events, with luminosities a significant fraction of that of their host galaxies during the peak of their explosions. Therefore they are relatively easy to detect at high redshift ($z\sim 1$). This specific type of supernova is characterized by the absence of a hydrogen line (or, indeed, a helium one) in their spectrum (which is typical of type I SNe) and instead by the presence of a singly-ionized silicon line at $615 \rm{nm}$, near peak light.

Despite a significant scatter, type Ia supernovae peak luminosities have been found to be very closely correlated with observed differences in the {\it shapes} of the light curves: dimmer SNe are found to decline more rapidly after maximum brightness, while brighter SNe decline more slowly~\cite{Phillips:1993ng,Riess:1996pa,Hamuy:1996sq}. This difference seems to be traceable to the amount of $^{56}\rm{Ni}$ produced in the explosion; more nickel implies both a higher peak luminosity and a higher temperature, and thus opacity, leading to a slower decline. 

After one adjusts for the difference in the light-curves, the scatter in peak luminosity can be reduced to $~ 15\%$. In this sense, SNeIa are referred to as ``standardizable candles" and are very good candidates for distance indicators, since one would expect any remaining difference in their peak luminosity to be due to a difference in distance.  Another important aspect of this uniformity is that it provides standard spectral and light curve templates that offer the possibility of singling out those SNe that deviate slightly from the norm~\cite{Perlmutter:1999rr}. 

After the necessary adjustments, all SNeIa have the same absolute magnitude $M$, and any difference in their apparent magnitude $m$ is attributable to a difference in their distance. The apparent and intrinsic magnitudes are related via the luminosity distance $d_L$
\be\label{app_vs_intr}
m=M+5\log{\left(\f{d_L}{10{\rm pc}}\right)}+K\,,
\ee
where $K$ is a correction for the shifting of the spectrum~\cite{Riess:1998cb,Perlmutter:1998np,Phillips:1993ng,Hamuy:1996sq,Nugent:2002si,Perlmutter:2003kf} necessary because only a part of the spectrum emitted is actually observed. In an expanding universe, the expression of the luminosity distance as a function of redshift is
\be\label{lum_distance}
d_L(z)=(1+z)\cdot\int_0^z\f{dz'}{H(z')}\,.
\ee
By independently measuring $d_L(z)$ and $z$, one can then constrain the expansion history $H(z)$. Following pioneering work reported in~\cite{Hansen}, the High-Z Supernova team~\cite{Riess:1998cb} and the Supernova Cosmology Project~\cite{Perlmutter:1998np} measured the apparent magnitudes of many SNeIa (of redshift $z \lesssim 1$) and directly determined their distances. They then compared these distances to those inferred from the redshifts of the host galaxies (measured from the spectra of the galaxies when possible, otherwise the spectra of the SNe themselves). The most distant SNe appeared dimmer than expected in a universe currently dominated by matter, and a careful comparison with low redshift supernovae allowed one to rule out the possibility that this dimming is due to intervening dust. Assuming a flat universe described by GR, and homogeneous and isotropic on large scales, the data are best fit by a universe which has recently entered a phase of accelerated expansion, i.e. a universe for which the deceleration parameter~(\ref{decelerationparameter}) is currently negative ($q_0 \simeq -1$). The simplest model fitting the data is a universe in which matter accounts for only about a quarter of the critical density, while the remaining $70\%$ of the energy density is in the cosmological constant $\Lambda$. This model is commonly referred to as the $\Lambda$CDM model. 
In fact, the SNe data by themselves allow a range of possible values for the matter and cosmological constant density parameters ($\Omega^0_{\rm M}$ and $\Omega^0_{\Lambda}$ respectively). 
\begin{figure}[t]
\begin{centering}
\includegraphics[width=120mm]{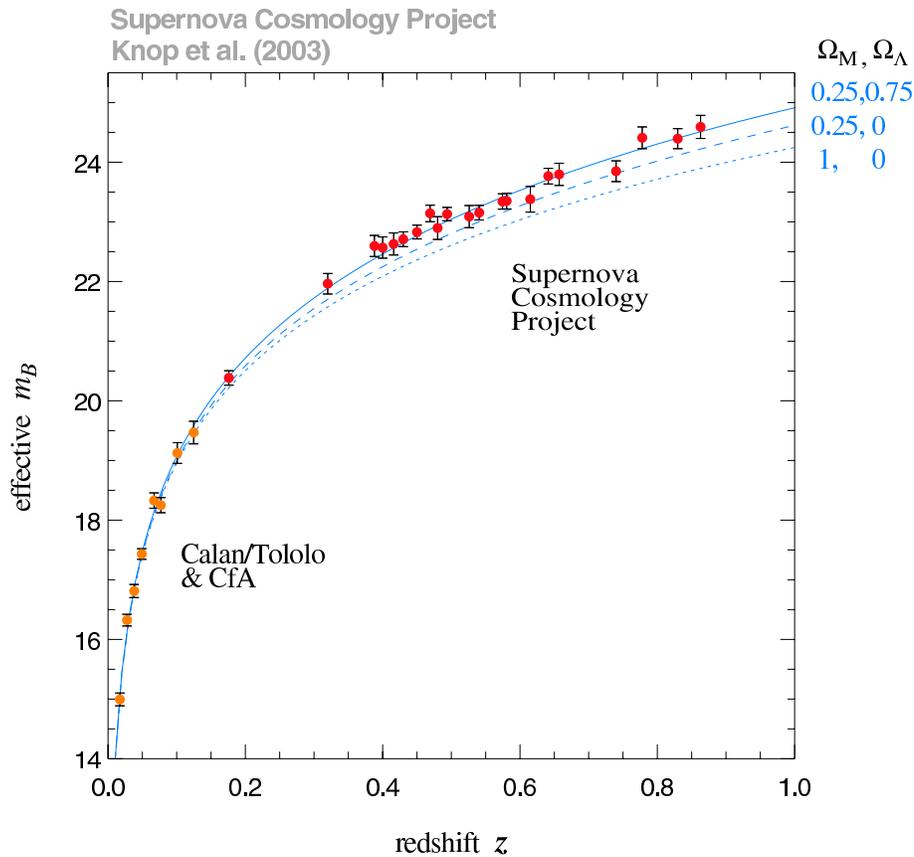}
\caption{Hubble diagram for Type Ia Supernovae, plotting the effective magnitude $m_B$ versus redshift $z$ (Knop {\it et al}.~\cite{Knop:2003iy}). The solid line is the best-fit cosmology and corresponds to a universe with $\Omega_{\rm M}^0=0.25$ and $\Omega_{\Lambda}^0=0.75$.}
\label{SNeIa}
\end{centering}
\end{figure}
However, as we will discuss in the following subsections, we can use what we know from other observations, such as the Cosmic Microwave Background (CMB)  and the large scale structure (LSS), to further constrain these parameters.

In more recent years, distant SNe with redshifts up to $z\sim 1.7$ have been observed by the Hubble Space Telescope (HST)~\cite{Astier:2005qq,Riess:2006fw}. These observations show that the trend toward fainter SNe at moderate redshifts has reversed; therefore they play a key role in disregarding dust as a plausible explanation of the dimming of intermediate SNe, and in firmly establishing~\cite{Starkman:1999pg,Huterer:2002wf} acceleration. More generally, SNe at lower and higher redshift are both important in the study of the expansion history. Nearby SNe ($z\lesssim 0.3$) are useful for defining characteristics of Type Ia SNe, understanding the explosion and exploring systematics; all combined, they are important for establishing distance indicators and are currently~\cite{Hicken:2009dk,Hicken:2009df} being measured, with more to come in the future~\cite{WoodVasey:2004pj,Freedman:2004uz}. Intermediate redshift SNe ($0.3\lesssim z \lesssim 0.8$) measure the strength of cosmic acceleration and are being measured by the ESSENCE project~\cite{WoodVasey:2007jb}, the SuperNova Legacy Survey (SNLS)~\cite{PalanqueDelabrouille:2005yf} and the Sloan Digital Sky Survey-II (SDSS-II) Supernova Survey~\cite{Frieman:2007mr}. Finally, distant SNe ($z\gtrsim 0.8$) are important to break the degeneracy among the cosmological parameters and they are currently being measured by the Higher-z Supernova Search Team (HZT)~\cite{Strolger:2004kk} and the GOODS team~\cite{:2003ig}.

In the near future, the space-based Joint Dark Energy Mission (JDEM)~\cite{JDEM}, (a wide-field optical-infrared telescope), will offer precision measurements with the aim of determining the nature of cosmic acceleration. This mission may incorporate the Baryon Acoustic Oscillation (BAO), Supernovae (SNe) and Weak Lensing (WL) techniques.
Also, upcoming and future surveys such as the earth-based Dark Energy Survey (DES)~\cite{DES} and the Large Synoptic Survey Telescope (LSST)~\cite{LSST} will provide observations useful for constraining the expansion history of the universe.

\subsection{Complementary Probes}
Supernovae are the first and most direct evidence for the late-time acceleration of the universe. However, independent evidence comes from complementary probes. Data from different observations are important not only to confirm the results from SNe, but also to help break some of the degeneracy in the cosmological parameters, such as the density parameters $\Omega^0_{\rm M}$ and $\Omega^0_{\Lambda}$. As we have mentioned, the measurement of the expansion history via SNe is an example of the use of standard candles to infer the luminosity distance as a function of the redshift $d_L(z)$. Alternatively, one may employ standard rulers, extracting the angular diameter distance $d_A(z)$ (given by the ratio of the object's actual size to its angular size) and then $H(z)$. (Both methods are commonly referred to as geometrical probes). The expansion history can further be constrained with the growth of structure, as we will see shortly. Finally, some methods are based on inferring the age of the universe from stellar ages. In what follows we review several complementary probes of the expansion history of the universe.

The CMB is the radiation that reaches us from the surface of last scattering -- that is, from the time at which photons decoupled from the thermal bath and started to free stream to us. It is in the form of an almost isotropic blackbody spectrum, with a temperature of approximately $2.7$ K. The observed blackbody distribution, typical of sources that are in thermal equilibrium, is strong evidence for the theory of the Big Bang. At early times, when the universe was radiation dominated, photons were easily energetic enough to ionize hydrogen atoms. Therefore, the universe was filled with a charged plasma, the {\it photon-baryon plasma}, and was opaque. As the universe expanded, the temperature decreased and the photons redshifted, losing energy until they were no longer able to ionize hydrogen. At T $\sim O(1)$ eV, corresponding to a redshift of $z\sim 1100$, neutral hydrogen formed and the photons decoupled from the plasma. At early times, the number and energy densities of matter were high enough that the plasma constituents were in thermal equilibrium, ensuring a blackbody spectrum for the radiation distribution. This spectrum is maintained by photons after decoupling, just at lower and lower temperatures. Indeed, the effect of homogeneous and isotropic expansion on a blackbody distribution is that of preserving the blackbody structure with temperatures that redshift with the wavelength, $T\propto 1/a$.

\begin{figure}[t]
\begin{centering}
\includegraphics[width=120mm]{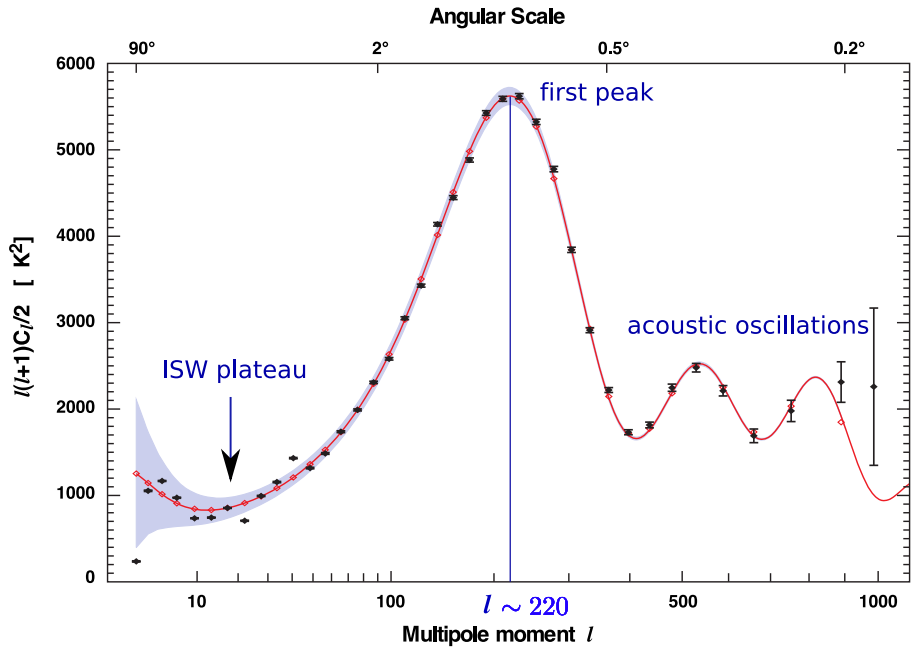}
\caption{The Cosmic Microwave Background angular power spectrum, representing the two-point correlation 
function of temperature fluctuations (in momentum space) on the sky as a function of angular separation, [courtesy of NASA/WMAP Science team]. The solid line is the best fit $\Lambda$CDM model. On larger scales, i.e. at lower multipoles {\it l}, we notice the Integrated Sachs-Wolfe plateau, at $l\sim220$ we see the first acoustic peak and on smaller angular scales the subsequent acoustic peaks}
\label{CMB_power}
\end{centering}
\end{figure}
As measured by the Cosmic Background Explorer (COBE)~\cite{Smoot:1992td}, the temperature distribution of the photons is not completely isotropic, with fluctuations of one part in $10^5$. These fluctuations contain a great deal of physical information about our universe, and the ongoing measurements from the Wilkinson Microwave Anisotropy Probe (WMAP)~\cite{Spergel:2003cb,Spergel:2006hy,Komatsu:2008hk} have reached unprecedented precision in using it to constrain cosmological parameters. It is worth exploring in a little more detail how this is achieved. 

Matter in the universe today is highly inhomogeneous and clustered. Our understanding of how this large scale structure developed is that initially small density perturbations, in an otherwise homogeneous universe, grew through gravitational instability into the objects we observe today. The basic observable in CMB physics is the temperature fluctuation on the sky, for which we measure the 2-point correlation function. The Fourier counterpart of the correlation function, the {\it power spectrum}, is what is commonly shown in plots. As can be seen in Fig.~\ref{CMB_power} , there are two major components of a typical CMB power spectrum:  the Sachs-Wolfe plateau at large scales and a series of peaks and troughs at smaller scales. These both carry important information about the source of cosmic acceleration. The fluctuations we observe on the largest scales are essentially the primordial ones, since the largest scales have only recently entered the horizon and therefore little causal physics has affected them. On smaller scales we instead observe fluctuations which entered the horizon before decoupling and have undergone {\it acoustic oscillations} characteristic of the photon-baryon fluid. 
\begin{figure}[t]
\begin{centering}
\includegraphics[width=80mm]{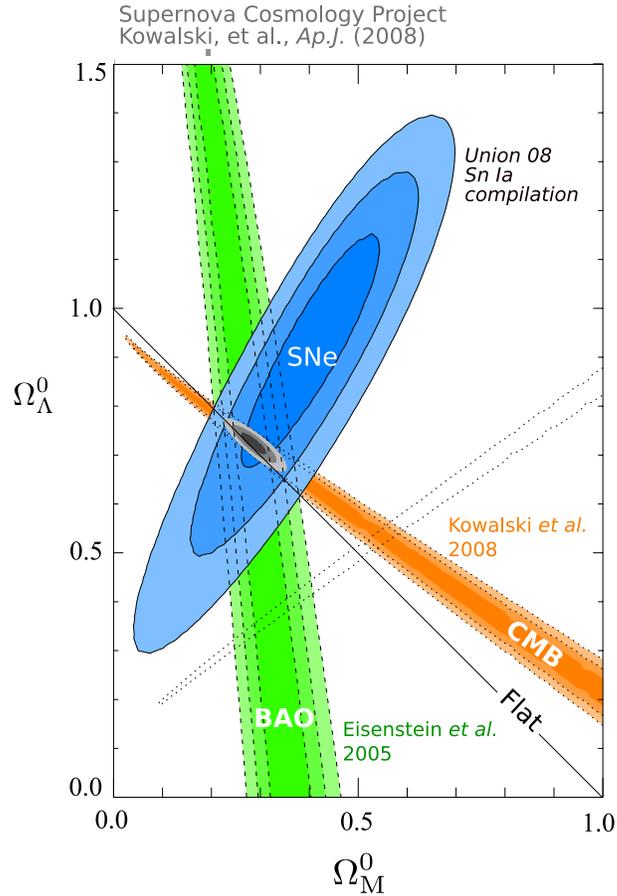}
\caption{Observational constraints in the $\Omega^0_{\rm M}-\Omega^0_{\Lambda}$ plane~\cite{Kowalski:2008ez}. The contours represent the $68.3\%$, $95.4\%$ and $99.7\%$ confidence level from CMB, BAO and SNeIa and their combination, with different colors corresponding to different data sets. The vertical green contours, centered around $\Omega^0_{\rm M}\simeq 0.3$, represent BAO constraints from the SDSS galaxy survey~\cite{Eisenstein:2005su}; the orange narrow diagonal contours correspond to constraints from WMAP-5 year observations of the CMB anisotropies (including a prior on the Hubble parameter)~\cite{Dunkley:2008ie}. Finally the wide blue diagonal contours represent constraints from supernovae Union Set~\cite{Kowalski:2008ez} and the gray-scale contours the combined constraints from BAO, SNeIa and CMB.}
\label{combined}
\end{centering}
\end{figure}

Let us focus on the first peak of these oscillations.  The acoustic oscillations are damped oscillations of the photon-baryon plasma under the competing effects of gravity and radiation pressure. The maximum amplitude of the oscillations is expected on fluctuations that entered the horizon just before decoupling. To see this, consider an overdense region of size $R$, which therefore contracts under self-gravity over a timescale $R$ (since we have set $c=1$). If $R\gg H^{-1}_{ls}$ then the region will not have had time to collapse over the lifetime of the universe at last scattering. If $R\ll H^{-1}_{ls}$ then collapse will be well underway at last scattering, matter will have had time to fall into the resulting potential well and result in a rise in temperature. This, in turn, gives rise to a restoring force from photons which acts to damp out the inhomogeneity. Clearly, therefore, the maximum anisotropy will be on a scale which has had just enough time to equilibrate, i.e. $R\sim H^{-1}_{ls}$. This means that we expect to see a peak in the CMB power spectrum at an angular size corresponding to the horizon size at last scattering. Since we know the physical size of the horizon at last scattering, this provides us with a ruler on the sky. The corresponding angular size will then depend on the spatial geometry of the universe. For a flat universe ($k=0$, $\Omega^0_{\rm{tot}}=1$), we expect a peak at $l\simeq 220$. As can be seen in Fig.~\ref{CMB_power}, the first peak indeed appears at this angular scale, providing strong evidence for the spatial flatness of the universe. 
As represented in Fig.~\ref{combined}, the CMB strongly constrains the sum of the matter and dark energy densities, 
offering a probe which is complementary to the SNe data in the $\Omega^0_{\rm M}-\Omega^0_{\Lambda}$ space. 

However, this is just one piece of the valuable information that can be extracted from the CMB. In general, the CMB power spectrum provides high precision constraints on essentially all of the cosmological parameters~\cite{Komatsu:2008hk}. The overall CMB spectrum, as well as the matter power spectrum as measured from LSS, are best fit by models which undergo late-time accelerated expansion. 
\begin{figure}[t]
\begin{centering}
\includegraphics[width=120mm]{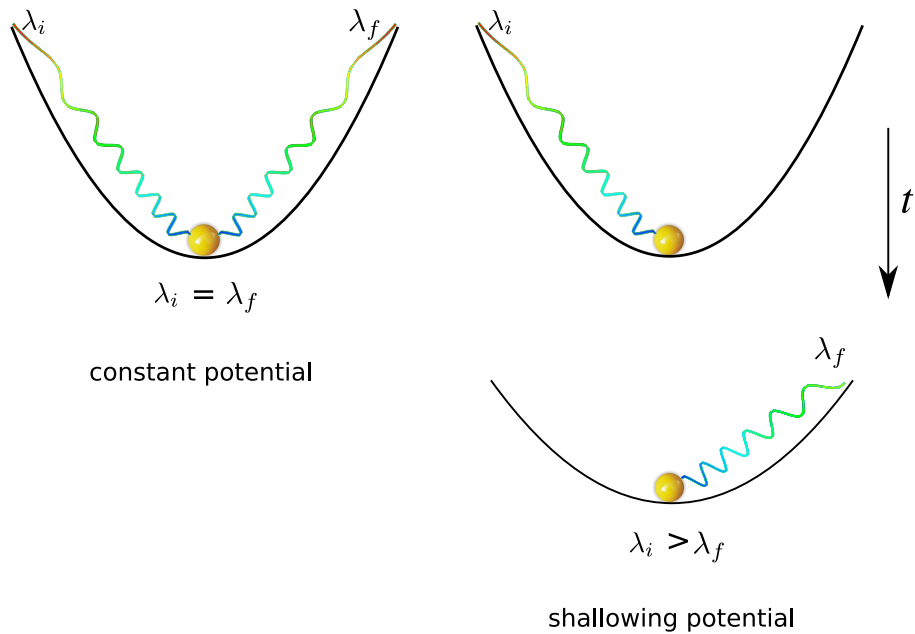}
\caption{Schematic representation of the Integrated Sachs-Wolfe effect (ISW) on CMB photons. On the left
side of the figure, a photon travels through a constant potential, gaining and losing the same amount of energy while passing through the potential well. In this case, the outgoing wavelength corresponds to the incoming one, and there is no net gain (or loss) of energy. On the right side of the figure, a photon travels through a shallowing potential. As the potential well from which the photon exits is shallower than the one it entered, there is an overall gain in energy and the outgoing wavelength is shorter than the incoming one.}
\label{ISW-schematic}
\end{centering}
\end{figure}

Information from the CMB concerning the total energy density of the universe can be combined with data about the distribution of matter, from which one extracts $\Omega^0_{\rm M}$. In addition, the CMB power spectrum itself offers a way to constrain the matter density via the spacing of the peaks of the acoustic oscillations. Complementary strong constraints on the matter density come from the observed distribution of matter in the universe. In particular, the turnover scale in the matter power spectrum is very sensitive to $\Omega^0_{\rm M}$ and the abundance of clusters of galaxies constrain the $\sigma_8$ parameter, which encodes information about $\Omega^0_{\rm M}$ and the fractional density perturbation.

The WMAP data together with LSS data from the Sloan Digital Sky Survey (SDSS)~\cite{SDSS} yield the value $\Omega^0_{\rm M}=0.25\pm0.03$ ($1\sigma$, in the context of a flat $\Lambda$CDM model)~\cite{Tegmark:2006az}. Combining these with the supernovae data, we can pin down a narrow region in the $\Omega^0_{\rm M}-\Omega^0_{\Lambda}$ space corresponding to $\Omega^0_{\rm M}\sim1-\Omega^0_{\Lambda}\sim0.27$, as shown in Fig.~\ref{combined}. 

Gravitational lensing - the distortion of light paths by the inhomogeneous metric around matter in the universe - is also an important complementary probe as it offers a direct handle on the mass distribution, without relying on bias factors to convert from visible to total matter. Of particular importance in this context is the {\it weak} gravitational lensing of distant galaxies by intervening foreground mass overdensities, dubbed {\it cosmic shear}. This offers a direct measurement of mass fluctuations in the universe and their redshift evolution. The statistic associated with cosmic shear, the shear angular power spectrum, is sensitive to dark energy via the expansion history as well as via the growth of structure.  From precise measurements of the cosmic shear and its cross-correlation with galaxies it is possible to extract geometrical probes of dark energy~\cite{Jain:2003tba,Bernstein:2003es} as well as an estimate of the matter density parameter~\cite{Fu:2007qq}. In section \ref{sec:tests} we will discuss in more detail the constraints on the growth of perturbations from weak lensing tomographic surveys.  

Further useful constraints on the matter density can be derived from measurements of the fraction $f_{\rm{gas}}$ of X-ray emitting gas to total mass in galaxy clusters. This ratio is a good indicator  of the overall baryon fraction in the universe, and depends on the distance to the cluster, offering a probe of the expansion history. The ratio of the gas (baryons) to total mass is found to be about $11\%$, which yields a cosmic baryon fraction of about $15\%$~\cite{Allen:2004cd,Allen:2007ue}, after estimated corrections. Combining these measurements with the baryon density inferred very precisely from Big Bang Nucleosynthesis, one can thus derive an estimate of the matter density parameter. 
\begin{figure}[t]
\begin{centering}
\includegraphics[width=80mm]{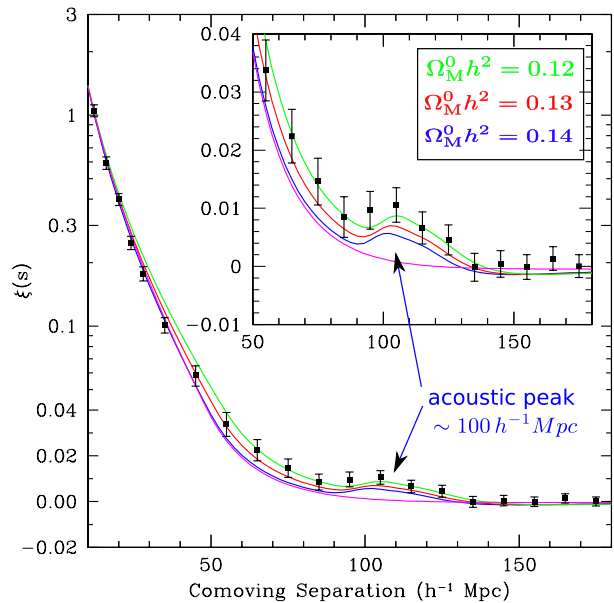}
\caption{The large scale redshift-space correlation function of the SDSS LRG (luminous red galaxies) sample, from~\cite{Eisenstein:2005su}. The magenta line shows a pure CDM  model ($\Omega_{\rm M}^0h^2=0.105$) which lacks the acoustic peak. The green, red and blue lines represent models with a baryon fractional density
of $\Omega_b^0h^2=0.024$ and a CDM density of, respectively, $\Omega_{\rm M}^0h^2=0.12,0.13,0.14$. The bump at $100\,h^{-1}{\textrm Mpc}$ corresponds to the acoustic peak. Details about the data set and error bars can be found in~\cite{Eisenstein:2005su}.}
\label{matter_power}
\end{centering}
\end{figure}

The large scale plateau of the CMB power spectrum offers a useful probe of the expansion history. In a matter dominated universe, when the largest fluctuations enter the horizon, the gravitational potentials are constant, and therefore do not imprint any additional anisotropy on the photons that traverse them. Indeed, a photon entering a potential well gains energy, while a photon climbing out of a potential loses energy, as represented in Fig.~\ref{ISW-schematic}; if the potential is constant in time, there is no overall change in the energy of the photon. However, if matter is not dominant and the late universe is undergoing a phase of accelerated expansion, then the gravitational potentials are expected to evolve. As a consequence, photons traveling through a shallowing (deepening) potential well will be blueshifted (redshifted), as they gain more energy entering the well than the energy they lose climbing out. This is known as the Integrated Sachs-Wolfe effect (ISW). However, this is a small contribution to the overall CMB power spectrum on the largest scales, approximately $10\%$ of the total signal, and therefore is very hard to extract. One proceeds by cross-correlating the CMB power spectrum with the galaxy one, as the ISW signal is expected to be strongly correlated with the distribution of matter around us. At present, the ISW effect has been measured out to redshift $z\sim 1.5$~\cite{Ho:2008bz,Giannantonio:2008zi} at a $4\,\sigma$ significance level, and constitutes a promising probe of dark energy that is likely to play an important role in distinguishing among candidate models of cosmic acceleration, as we will discuss in section~\ref{sec:MGR}. However, it must be noted that the ISW effect will always be limited by a low signal to noise ratio.

The acoustic oscillations that we observe in the spectrum of the CMB also leave a characteristic imprint on the matter power spectrum~\cite{Peebles:1970ag,Sunyaev:1970er,Bond:1984fp,Holtzman:1989ki}, commonly referred to as Baryon Acoustic Oscillations (BAO). These imprints provide a new opportunity to measure distances at different redshifts. Indeed, the length scale that was imprinted by the sound waves, referred to as the {\it acoustic scale}, persists to the present time in the clustering of matter at a characteristic comoving length of $~100\,h^{-1}\,{\textrm{Mpc}}$. The latest generation of galaxy redshift surveys is able to probe the length scales required to make a precision measurement of the BAO signal, and this method has therefore been proposed as a new geometrical probe to constrain the physics of cosmic acceleration~\cite{Hu:1996vq,Eisenstein:1997ik,Blake:2003rh, Hu:2003ti,Seo:2003pu}. 
The acoustic scale corresponds to a preferred length scale on the surface of last scattering, i.e. the distance that a sound wave propagating from a point source at the end of inflation would have traveled before decoupling. In a redshift survey, this length scale can be measured both along and across the line of sight, allowing the observer to reconstruct the Hubble parameter and the angular diameter distance as a function of redshift, respectively $H(z)$ and $d_A(z)$.
The BAO have been detected in the SDSS data~\cite{Eisenstein:2005su} data at a redshift of approximately $z\sim0.35$ and in a combination of the Two Degree Field Galaxy Redshift Survey (2dFGRS) and SDSS data~\cite{Percival:2007yw}. 

From the recent WMAP 5-year results~\cite{Komatsu:2008hk}, the combined constraints from BAO, CMB and SNeIa on the density parameters are: $\Omega^0_b =0.0456\pm0.0015$, $\Omega^0_{\rm M} h^2=0.1358^{+0.0037}_{-0.0036}$ and $\Omega^0_{\Lambda}=0.725\pm0.015$ to $68\%$ C.L..


\section{The Cosmological Constant}
\label{sec:cosmoconstant}

The natural tendency of a universe filled with regular matter is to decelerate, as the mutual gravitational attraction fights against any initial expansion. This was clear even to Einstein, for whom this fact was an obstacle to obtaining a static universe, rather than an accelerating one. Indeed, as Einstein realized almost immediately, there is an extension of the field equations of general relativity, consistent with general covariance and second order field equations, that can battle contraction and ultimately lead to acceleration - the cosmological constant.

The cosmological constant $\Lambda$ is the coefficient of the term in the field equations that is proportional to the metric. It appears in the action for gravity as
\be
S=\frac{\mpl^2}{2}\int d^4x \sqrt{-g} \left(R-2\Lambda\right) \ ,
\ee
and in the resulting Einstein field equations as
\be
R_{\mu\nu}-\frac{1}{2}Rg_{\mu\nu} +\Lambda g_{\mu\nu}=\frac{1}{\mpl^2} T_{\mu\nu} \ ,
\ee
where $T_{\mu\nu}$ is the energy-momentum tensor for mass-energy sources in the universe.

As we have mentioned earlier, for cosmological purposes, we may treat the cosmological constant as a component of the energy density in the universe, since adding a cosmological constant to the Einstein equation is equivalent to including an energy-momentum tensor of a perfect fluid with equation-of-state parameter is $ w_{\Lambda}=-1$. 

As we have also seen, the cosmological constant is a prime candidate for the agent causing cosmic acceleration and, subject to ever more accurate data, provides a rather good fit to all known cosmological observations. Furthermore, in classical general relativity the 
cosmological constant is merely a constant of nature, with dimensions of [length]$^{-2}$, in the same sense that the Planck mass in nothing but a dimensionful constant of the theory.  In this setting, it is meaningless to enquire about the value of the constant, rather it is just something that we determine through experiment. However, when we begin to think about the quantum nature of matter it becomes clear that the cosmological constant poses a deep puzzle for fundamental physics.

Particle physics, described by quantum field theory, is a remarkably accurate description of nature, and one of the best tested theories in all of science. Within this theory, we can attempt to calculate the expected contribution to the cosmological constant from 
quantum fluctuations in the vacuum.  Taking the Fourier transform of a free (noninteracting) quantum field, each mode of fixed wavelength is a simple
harmonic oscillator, which we know has zero-point (ground state) energy
$E_0 = {1\over 2}\hbar \omega$, where $\omega$ is the frequency of the oscillator.  The net energy of these modes is then
given by a divergent integral. Thus, our most naive expectation of the vacuum energy calculated within field theory is that it should be infinite! Perhaps, however, we have been too hasty in performing this calculation. After all, we have included modes of arbitrarily small wavelength, and this presumes that we can trust our descriptions of matter and gravity down to such tiny scales. Without a well-understood and calculable quantum theory of gravity, this is surely not reasonable.
A better estimate of the vacuum energy can therefore be obtained by introducing a cutoff energy, above which we ignore the contributions of any modes, assuming that this will be justified by a more complete theory. This approach corresponds to assuming that effective field theory correctly describe the appropriate limit of quantum
gravity.  In that sense, if a parameter has dimensions [mass]$^n$, then we expect the corresponding mass parameter
to be driven up to the scale at which the effective description
breaks down.  Thus, if classical
general relativity is valid up to the Planck scale, we expect the
vacuum energy to be given by
\be
  \rho_{\rm vac}^{\rm (theory)} \sim M_{\rm P}^4 \ .
  \label{rhoguess}
\ee

Comparing this value to the value $\rho_{\rm vac}^{\rm (obs)}$ required to explain cosmic acceleration, we obtain 
\be
  \rho_{\rm vac}^{\rm (obs)}
  \sim  10^{-120}\rho_{\rm vac}^{\rm (theory)}\ ,
  \label{rhoobs}
\ee
which is embarrassing, to say the least. It is perhaps more sensible to express the vacuum energy in
terms of a mass scale as $ \rho_{\rm vac} = M_{\rm vac}^4$, in terms of which the value required to explain our observations is 
$M_{\rm vac}^{{\rm (obs)}} \sim 10^{-3}{\rm ~eV}$, satisfying
\be
  M_{\rm vac}^{{\rm (obs)}} \sim 10^{-30} M_{\rm vac}^{{\rm (theory)}}\ .
  \label{mvac1}
\ee
Nevertheless, this discrepancy of 30 orders of magnitude in energy is overwhelming, and is what is meant by
{\it the cosmological constant problem}.  

One may add to this problem the following puzzling observation. The ratio of the vacuum
and matter densities changes as the universe expands according to
\be
  {\Omega_\Lambda \over \Omega_{\rm M}} = {\rho_\Lambda
  \over \rho_{\rm M}} \propto a^3\ .
\ee
Thus, only during a brief epoch of cosmic history is it possible for observers to
witness the transition from matter domination to $\Lambda$ domination, during which $\Omega_{\Lambda}$ and $\Omega_{\rm M}$ are of the same order of magnitude. This is known as {\it the coincidence problem}.

The issue of reliably calculating the cosmological constant, and finding a framework in which that calculation leads
to a result dramatically different to the expected one has proven remarkably resistant to theoretical attack. It is fair to say that
there are not currently any especially promising approaches. Nevertheless, there are two theoretical lines of research that are worth mentioning in this context.

The first is supersymmetry (SUSY). Supersymmetry is a spacetime symmetry that relates particles of different spins (there exists a matching fermionic degree of freedom for every bosonic one, and vice-versa.) Since bosons and fermions make opposite sign contributions to the vacuum energy, in a truly supersymmetric  theory the net vacuum energy should be zero, with cancellations occurring between, for example, the spin-1/2 electron and its spin-0 partner, the ``selectron'' (with the same mass and charge as the electron.) 

It is, however, entirely obvious from experiment and observation, that the world is not supersymmetric - we would certainly know if
there existed a scalar particle with the same mass and charge as an electron. Thus, if supersymmetry is realized in nature, it must
be broken at some scale $M_{\rm susy}$ which current bounds force to be greater than around a TeV.  An analogous calculation of the vacuum energy in such a theory then yields
\be
  M_{\rm vac} \sim M_{\rm susy}\ ,
\ee
with $\rho_{\rm vac} = M_{\rm vac}^4$.  That SUSY is a beautiful idea provides us with no clue as to the possible value of
$M_{\rm susy}$. However, independent of the cosmological constant problem, or any cosmological data, SUSY has been suggested as a solution to one of the outstanding problems of the standard model of particle physics - the hierarchy problem: why is the scale of electroweak symmetry breaking so much smaller than the scale of gravity. SUSY can help with this problem if the SUSY breaking scale is very close to current bounds, i.e.
\be
  M_{\rm susy} \sim 10^3{\rm ~GeV}\ ,
\ee
leading to the hope that we may find evidence for SUSY at the upcoming Large Hadron Collider (LHC). However, even this
lower bound on the SUSY breaking scale leaves us with
\be
  M_{\rm vac}^{\rm (obs)} \sim 10^{-15}M_{\rm susy}\ ,
  \label{mvac2}
\ee
which is still nowhere near acceptable.

Nevertheless, should we find evidence for supersymmetry through particle physics experiments, the question of the
strangely small value of the cosmological constant will become one of the SUSY breaking sector of the theory, and we will have made some progress, although how much is unclear.

A second theoretical approach to the cosmological constant and cosmic acceleration problems is to consider the logical possibility
that the vacuum energy is a feature of our local environment - {\it the anthropic principle}. This is an idea that may make logical sense if there exist many different possible values of physical parameters (and the cosmological constant in particular), and a mechanism in the early universe for sampling these possibilities, either over and over again in time, or in different regions of space.

Clearly, only a certain range of values of the cosmological constant is consistent with the development of life like us (and there are a number of assumptions here), since if $|\Lambda |$ is significantly larger than that required to yield the observed cosmic acceleration, then the universe would have recollapsed (if $\rho_{\rm vac}$ were negative), or
expanded so quickly that galaxies couldn't form (if $\rho_{\rm vac}$ were positive). Given this, some authors have argued~ \cite{Linde:1986fd,Weinberg:1987dv,Vilenkin:1994ua,Martel:1997vi,Garriga:1999bf,Garriga:2002tq} that, under an assumption about the distribution of possible values of $\rho_{\rm vac}$, we should expect a value of $\Lambda$ close to that we seem to have observed.

The recent resurgence of interest in this possibility stems from the realization that within string theory
there seems to exist a large number (more than $10^{100}$) of possible vacuum states~\cite{Dasgupta:1999ss,Bousso:2000xa,Feng:2000if,
Giddings:2001yu,Kachru:2003aw,Susskind:2003kw}. These vacua arise from different compactification possibilities, and the roles played in them by various brane and gauge field configurations. The low energy effective theories constructed around these metastable vacua should have different local values of the vacuum energy (and other physical parameters.) An approach to addressing the cosmological constant problem then begins by asserting that it is highly likely that there must exist (and, eventually, hopefully identifying) one or more vacua with a cosmological constant (and perhaps other parameters) that take the values one might think this reasonable given the number quoted above.

The second step in constructing an anthropic argument around the string theory landscape involves arguing that there is a mechanism to sample, or populate, the whole distribution of possible vacua, in order to be able to claim that if a phenomenologically viable vacuum exists then some part of the universe will exist in that state. The mechanism of choice in the string theory landscape is {\it eternal inflation}. Inflation itself works to expand a small region of space to a size larger than the observable universe and, in general, models of inflation tend to be eternal, meaning that the universe continues to inflate in
some regions even after inflation has ended in others~\cite{Linde:1986fd,Vilenkin:1983xq}. In this model many different domains may be inflated separately, and then small parts of the universe may cease to inflate and drop out into different final vacuum states. Thus, if there exist possible vacua with a cosmological constant of the right size to explain cosmic acceleration, parts of the universe that populate those vacua will cease inflating, reheat the universe, undergo standard cosmic evolution, and ultimately begin accelerating again at late times.

This is a brief description of how the anthropic principle may be realized in the context of the string theory landscape. It is however, too early to know whether it can be realistically implemented, and even harder to say whether one will ever be able to falsify, and hence make science out of, such an approach.

Nevertheless, this is the best developed argument regarding the possible small value of $\Lambda$, and perhaps for the value needed to explain cosmic acceleration.


\section{Dark Energy}
\label{sec:darkenergy}

As we discussed in section~\ref{sec:observations}, while the data clearly points to an accelerating universe, it does not single out a cosmological constant as the clear source of this acceleration. Let us assume, therefore, that the cosmological constant is zero, or at least so small as to be unable to account for the current period of accelerated expansion, and explore what other sources of mass-energy in the universe may be responsible (see~\cite{Bean:2005ru} for a discussion of logical possibilities, and~\cite{Copeland:2006wr} for a detailed review). Cosmologists have asked this question before, when confronted with the inflationary paradigm for solving the problems of early universe cosmology, and so we do not come the issue completely cold.

The most faithful servant of the theoretical cosmologist is the trusty scalar field. It has served us dutifully not only through inflation, but also baryogenesis models, topological defect formation and some dark matter models, and so it is only natural that we turn to it now, in our epoch of need. Consider a real scalar field $\phi$, with potential $V(\phi)$, with Lagrangian density
\be
{\cal L}= \frac{1}{2}g^{\mu\nu}\partial_{\mu}\phi\partial_{\nu}\phi -V(\phi) \ ,
\ee
for which the energy-momentum tensor is given by
\be
T_{\mu\nu}=\partial_{\mu}\phi\partial_{\nu}\phi -g_{\mu\nu}\left[\frac{1}{2}g^{\sigma\rho}\partial_{\sigma}\phi\partial_{\rho}\phi +V(\phi)\right] \ .
\label{scalarstress}
\ee
If the scalar field is homogeneously distributed in space, as we shall assume, then~(\ref{scalarstress}) takes the perfect fluid form, with energy density and pressure given by
\bea
\rho &=& \frac{1}{2}{\dot \phi}^2 + V(\phi) \nonumber \ , \\
p &=& \frac{1}{2}{\dot \phi}^2 - V(\phi) \nonumber \ ,
\eea
and we may then associate with it an instantaneous equation of state parameter
\be
w_{\phi}=\frac{{\dot \phi}^2 -2V(\phi)}{{\dot \phi}^2 +2V(\phi)} \ .
\ee

As we learned when studying inflation, in the slow-roll regime, in which the scalar field equation of motion
\be
{\ddot \phi} +3H{\dot \phi} + V_{, \phi} =0 \ ,
\ee
(where $V_{,\phi}\equiv\partial V/\partial\phi$) becomes ${\dot \phi}\simeq -\frac{1}{3H} V_{, \phi}$, the kinetic energy contributions to the pressure and the energy density become subdominant, and the equation of state parameter can become sufficiently negative to drive cosmic acceleration. 

Unlike inflation however, we need not require this period of acceleration to end, or the universe to be reheated after it. Rather, our requirements are that the energy scale associated with the potential be such that acceleration is happening at the current epoch, and that the energy density in the scalar field comprise the measured value of $\Omega_{\rm DE}$. Two of the most studied examples of this are inverse power law potentials
\be
V(\phi) =M^4\left(\frac{\phi}{M_p}\right)^{-n} \ ,
\ee
and exponential potentials
\be
V(\phi)=M^4\exp\left(-\alpha\frac{\phi}{M_p}\right) \ ,
\ee
where $M$ is some mass scale and $\alpha$ is a dimensionless parameter.

In the latter case, it is well known that in the limit that the scalar field becomes so dominant that one can neglect contributions from other sources (such as matter), the scalar field and Friedmann equations can be solved exactly, to yield
\bea
\phi(t) &=& \frac{2}{\alpha}\ln\left(\frac{M^2\alpha^2}{\sqrt{2(6-\alpha^2)}}\frac{t}{M_p}\right) \nonumber \\
a(t) &=& a_0 t^{2/\alpha^2} \ . \nonumber
\eea
Therefore, acceleration (${\ddot a}>0$) is achieved if $\alpha^2<2$.

Including matter changes this conclusion of course. One effect is that the actual functional form of the accelerating solution is altered from the simple power law one. However, a more interesting modification is the possible existence of {\it tracker} solutions (in which the energy density in the scalar field approaches a fixed fraction of that in the other components, rather insensitively to the initial conditions), which provide a way to ameliorate the coincidence problem of cosmic acceleration.

While these simple models provide a basic framework in which to obtain cosmic acceleration from a dynamical component, they are not without problems. Perhaps the largest issue is the extreme fine-tuning imposed for the same reasons as we described when discussing the cosmological constant. It is one thing to write down a potential, but quite another to provide a reason for the energy scales involved, and an argument why they should remain stable in any sensible particle physics implementation of the idea.

To understand the fine-tuning, consider the mass of the quintessence field by expanding the potential about a point as $V(\phi)=(1/2)m_{\phi}^2\phi^2$. As we discussed in the previous section, to match observations, the energy density in whatever component is responsible for cosmic acceleration must be $\sim (10^{-12}\ {\rm GeV})^4$. Assuming slow-roll, and that $\langle\phi\rangle\sim\mpl$, we may then equate these two to obtain
\be
m_{\phi} \sim 10^{-33}\ {\rm eV} \ .
\ee
Obtaining this kind of value from a particle physics model is an extreme challenge, consisting of two parts. First, it is not a simple matter to construct a model in which such low energy scales arise in the first place. Second, we face the issue confronting all scalar fields - how are we to protect an unnaturally small choice of mass scale against quadratic divergences when we compute quantum corrections to the couplings?

As with inflation, one possibility is to choose a quintessence model in which a small mass is protected as a small parameter describing deviations from a point of enhanced symmetry in model space. This can be implemented~\cite{Frieman:1995pm} by making the quintessence field a Pseudo-Nambu-Goldstone Boson (PNGB) of a spontaneously broken shift symmetry. In the absence of soft breaking, the potential for such a field is a precisely flat function of an angular coordinate. Thus, if the potential is lifted by a small mass term the exact symmetry keeping the potential flat is only softly broken, and the potential takes the form
\be
V(\phi)=M^4\left[1+\cos\left(\frac{\phi}{f}\right)\right] \ ,
\ee
where $M$ is the overall mass scale, and $f\sim \mpl$ measures the range of variation of $\phi$. Expanding this potential around $\phi =0$ yields a field with a small mass (for appropriately small $M$), which is protected from quadratic divergences since the exact $U(1)$ axion symmetry is restored in the $f\rightarrow\infty$ limit.

We shall not dwell on the simplest quintessence models in this review, since space is limited and our intention is to provide some idea as to more recent work in the field. The model we have presented above admits a number of generalizations which have their own interesting features and signatures. 

\subsection{Phantom Models}
As we have described earlier, observations allow for the possibility that the equation of state parameter for dark energy obeys $w<-1$. However, thus far the models we have constructed do not venture into this parameter region. There is a good reason for this: in general relativity, it is conventional to restrict the possible energy-momentum tensors by imposing {\it energy conditions} in order to prevent instabilities of the vacuum or the propagation of energy outside
the light cone.  Standard examples, where in parentheses we have evaluated the condition for a perfect fluid source, are
\begin{itemize}
\item The Weak Energy Condition: $T_{\mu\nu} t^\mu t^\nu \geq 0$ $\forall$ timelike $t^\mu$ ($\rho \geq 0$
  and $\rho + p \geq 0$.)
\item The Null Energy Condition: $T_{\mu\nu} \ell^\mu \ell^\nu \geq 0$ $\forall$ null $\ell^\mu$ ($\rho + p \geq 0$.)
\item The Dominant Energy Condition: $T_{\mu\nu} t^\mu t^\nu \geq 0$ $\forall$ timelike $t^\mu$, and $T_{\mu\nu} T^\nu{}_\lambda t^\mu t^\lambda \leq 0$ ($\rho \geq |p|$.)
\item The Null Dominant Energy Condition: $T_{\mu\nu} \ell^\mu \ell^\nu \geq 0$ $\forall$ null $\ell^\mu$ and $T^{\mu\nu} \ell_\mu$ is a non-spacelike vector (as for the dominant energy condition, except that negative densities are allowed if $p=-\rho$.)
\item The Strong Energy Condition: $T_{\mu\nu} t^\mu t^\nu \geq {1\over 2}T^\lambda{}_\lambda
  t^\sigma t_\sigma$ $\forall$ timelike $t^\mu$ ($\rho + p \geq 0$ and $\rho + 3p \geq 0$.)
\end{itemize}

Since we are interested in an energy condition that implies stability, while allowing for the possibility of a cosmological constant, Garnavich~\cite{Garnavich:1998th} has advocated the use of the null dominant energy condition, boiling down to demanding that dynamical fields obey the dominant energy condition, while making an exception for the cosmological constant. Nevertheless, given what is allowed by the data, we may ask what happens if we allow for dark energy that violates the null dominant energy condition; i.e. obeys $w < -1$.

The possibility of such {\it phantom} components has been suggested by many authors~\cite{Caldwell:1999ew,Sahni:1999gb,Parker:1999td,Chiba:1999ka,Boisseau:2000pr,Schulz:2001yx,Faraoni:2001tq,Maor:2001ku,Onemli:2002hr,Torres:2002pe,Frampton:2002tu}, and the simplest example~\cite{Caldwell:1999ew} is provided by a scalar field $\chi$ 
with {\it negative} kinetic and gradient energy
\be
  \rho_\chi = -{1\over 2}{\dot\chi}^2 - {1\over 2}(\nabla\chi)^2
  + V(\chi)\ .
\ee
The associated equation of state parameter is
\be
  w={p_\chi \over \rho_\chi} =
  -\frac{V(\chi)+{1\over 2}\left[\dot{\chi}^2+a^{-2}(\nabla\chi)^2\right]}
  {V(\chi)-{1\over 2}\left[\dot{\chi}^2+a^{-2}(\nabla\chi)^2\right]} \ ,
\ee
obeying $w\leq -1$.

While we will not describe the behavior of phantom fields in any detail, it is worth pointing out that, classically, the cosmology of such a system is relatively simple. For example, if the
universe contains only dust and phantom matter. Then, if matter domination ceases at cosmological time $t_m$, then the
solution for the scale factor is
\begin{equation}
\label{constwscalefactor}
a(t)=a(t_m)\left[-w+(1+w)\left(\frac{t}{t_m}\right)\right]^{2/3(1+w)} \ ,
\end{equation}
from which it is easy to see that phantom matter eventually comes to dominate 
the universe. Since the Ricci scalar is given by
\begin{equation}
\label{wconstricciscalar}
R=\frac{4(1-3w)}{3(1+w)^2}\left[t-
	\left(\frac{w}{1+w}\right)t_m\right]^{-2} \ ,
\end{equation}
there is a future curvature singularity at $t=wt_m/(1+w)$ - the so-called {\it big rip}. 

It is, however, simple to construct models of phantom energy in which a
future singularity is avoided. With a suitable choice of potential, we may construct a simple
scalar field model that exhibits a period of time in which the expansion
proceeds with $w<-1$ and yet settles back to $w\geq -1$ at even later times, thus sidestepping the predictions of
$w=$~constant models, and avoiding the big rip.

However, the background cosmology aside, it is relatively straightforward to see that fluctuations in a phantom field
have a negative energy, and that it may therefore be possible
for the vacuum to decay into a collection of positive-energy 
and negative-energy particles.  Thus, if this instability is present on a timescale shorter than the age of the universe, then
such a phantom component cannot be a viable candidate for dark energy. Such a question has been studied in some detail by several authors~\cite{Carroll:2003st,Cline:2003gs}. For example, in~\cite{Carroll:2003st}, the authors considered a specific 
toy model in which the null dominant energy condition is violated, but in which the cosmology is well-
behaved and the theory is stable to small perturbations. In many ways, this model has been given the maximum opportunity to avoid a problem with instabilities. Nevertheless, what is relevant is a field theory calculation of the decay rate of phantom particles into
gravitons. Clearly, this decay rate would be infinite if the phantom theory was fundamental, valid up to arbitrarily high momenta, and would render the theory useless as a dark energy candidate. Considering instead the phantom theory to be an effective theory valid below a scale $\Lambda$, and including in the Lagrangian operators of all possible dimensions, suppressed by suitable powers of the cutoff scale, it was found that such higher order operators, even though they may be very high order, can lead to unacceptably short lifetimes for phantom particles unless the cutoff scale is less than $100$ MeV. This is, of course, difficult to reconcile with current accelerator experiments.

For this, and other reasons, it has proven extremely difficult to construct sensible models for which an equation of state $w<-1$ is realized within General Relativity with minimally-coupled sources, although some progress has been made with ghost condensate models~\cite{Creminelli:2008wc}. It is, however, important to emphasize that well-behaved energy-momentum sources can nevertheless yield an effective equation of state parameter that is less than $-1$ if the dark energy is non-minimally coupled~\cite{Maor:2001ku}, or if the underlying gravitational theory is not GR~\cite{Sahni:2002dx,Carroll:2004hc,Lue:2004za}.

\subsection{K-Essence}
In addition to considering the behavior of general potentials for scalar fields, it is also possible to think about the effects of modifying the kinetic term, yielding a class of models called {\it K-essence}~\cite{ArmendarizPicon:2000dh,ArmendarizPicon:2000ah,Mukhanov:2005bu,Vikman:2006hk,Garriga:1999vw}. 

To simplify notation in doing this, let us define $X\equiv -\frac{1}{2}g^{\mu\nu}\partial_{\mu}\phi\partial_{\nu}\phi$. Thus, a general action for a scalar field takes the form.
\be
S=\int d^4x\sqrt{-g} L(\phi,X) \ ,
\ee
with $L$ an unspecified function, and the canonical scalar field action corresponding to the special case $L(\phi,X)=X-V(\phi)$. While not completely general, an interesting subset of these actions are those with Lagrangian of the form
\be
L(\phi,X)=K(X)-V(\phi) \ ,
\ee
with $K(X)$ a positive semi-definite kinetic function, to be specified. 

If we are interested in background cosmological solutions, with the FRW ansatz for the metric, we must assume a homogeneous form for $\phi$, yielding $X=\frac{1}{2}{\dot \phi}^2$. The energy-momentum tensor for this field is straightforward to calculate and yields the usual perfect fluid form with pressure
$p$ and energy density $\rho$ given by
\begin{equation}
\label{pressure}
p={\cal L}=K(X)-V(\phi) \ ,
\end{equation}
\begin{equation}
\rho=\left[2XK_{,X}(X)-K(X)\right]+V(\phi) \ ,
\end{equation}
where a subscript $,X$ denotes a partial derivative with respect to $X$ Then, defining $w_{\phi}\equiv p/\rho$ one obtains
\begin{equation}
\label{w}
w_{\phi}=\frac{K(X)-V(\phi)}{\left[2XK_{,X}(X)-K(X)\right]+ V(\phi)} \ .
\end{equation}
 
Requiring that the energy density of the theory satisfy $\rho>0$ yields
\begin{equation}
\label{positiveenergy}
g(X)-2Xg_{,X}(X)<\frac{V(\phi)}{f(\phi)} \ ,
\end{equation}
and a further necessary condition that the theory be stable is that the speed of sound of
$\phi$ be
positive~\cite{Garriga:1999vw}. This yields
\begin{equation}
\label{soundspeed}
c_s^2\equiv \frac{\partial p}{\partial \rho} =
\frac{p_{,X}}{\rho_{,X}}=\frac{K_{,X}(X)}{K_{,X}(X)+2XK_{,XX}(X)}>0 \ .
\end{equation}

Solving the equations of motion in concert with the resulting Einstein equation, it has been 
shown~\cite{ArmendarizPicon:2000dh,ArmendarizPicon:2000ah,Mukhanov:2005bu,Vikman:2006hk,Garriga:1999vw} that
{\it tracking} behavior can be obtained. This means that for a wide range of
initial conditions, the energy density of the field $\phi$ naturally evolves so
as to track the energy density in matter, providing some insight into why dark
energy domination began only recently in cosmic history. However, as in all rolling scalar models, 
some fine-tuning remains, since one must ensure the right amount of dark energy density today. Further, just as with
minimal models, it is possible~\cite{Melchiorri:2002ux} to obtain phantom-like behavior, but once again, this comes at the risk of serious instabilities.

Most interestingly though, the varying speed of sound (compared to $c_s^2=1$ for
canonical models) opens the window to possible observational signatures of these models (which can otherwise be difficult to distinguish from canonical ones~\cite{Malquarti:2003nn}) in observations of large scale structure
in the universe.

\subsection{Coupled Dark Energy Models}
Our cosmological models seem to require both dark matter and dark energy. A logical possibility is that these dark sectors interact with each other or with the normal matter~\cite{Damour:1990tw,Carroll:1998zi,Amendola:1999er,Sahni:1999qe,Ziaeepour:2000qq,Bean:2000zm,Bean:2001ys,Majerotto:2004ji,Das:2005yj,Lee:2006za,Kesden:2006zb} and, indeed, a number of models have been proposed that exploit this possibility to address, for example, the coincidence problem.

There are, of course, many ways in which couplings may arise, and it is not possible to provide a comprehensive treatment here. Instead, let us mention an interesting subclass of models, defined by the action
\bea
S[g_{ab},\phi,\Psi_{\rm j}]
&=& \int d^4x\sqrt{-g}
\left[ \frac{1}{2} \mpl^2 R
-\frac{1}{2} (\nabla \phi)^2
 - V(\phi)
\right] \nonumber \\
&&+ \Sigma_{\rm j} S_{\rm j}[e^{2 \alpha_{\rm j}(\phi)} g_{\mu\nu}, \Psi_{\rm j}] \ ,
\label{action0}
\eea
where $g_{\mu\nu}$ is the Einstein frame metric, $\phi$ is a scalar field
which acts as dark energy, and $\Psi_{\rm j}$ are the matter fields. The functions $\alpha_{\rm j}(\phi)$ are couplings to the j${}^{th}$ matter sector.

This general action encapsulates many models studied in the literature~\cite{Carroll:2003wy,Chiba:2003ir,Damour:1990tw,Khoury:2003aq,Khoury:2003rn,Brax:2004qh,Brax:2004px,Brax:2005ew}.
The field equations are
\bea
\mpl^2 G_{\mu\nu} &=& \nabla_\mu \phi \nabla_\nu \phi - \frac{1}{2} g_{\mu\nu}
(\nabla \phi)^2 - V(\phi) g_{\mu\nu} \nonumber \\
&& + \sum_{\rm j} e^{4 \alpha_{\rm j}(\phi)}
\left[ ({\bar \rho}_{\rm j} + {\bar p}_{\rm j}) u_{{\rm j}\,\mu} u_{{\rm j}\,\nu} + {\bar p}_{\rm j} g_{\mu\nu} \right] \ ,
\label{ee0d}
\eea
\be
\nabla_\mu \nabla^\mu \phi - V'(\phi) = \sum_{\rm j} \alpha^{\rm j}_{,\phi}(\phi) e^{4 \alpha_{\rm j}(\phi)}
({\bar \rho}_{\rm j} - 3 {\bar p}_{\rm j} ) \ ,
\label{eq:scalar10a}
\ee
where we have treated the matter field(s) in the j${}^{th}$ sector as a fluid with density
${\bar \rho}_{\rm j}$ and pressure ${\bar p}_{\rm j}$ as measured in
the frame $e^{2 \alpha_{\rm j}} g_{\mu\nu}$, and with
4-velocity $u_{{\rm j}\,\mu}$ normalized according to $g^{\mu\nu} u_{{\rm j}\,\mu} u_{{\rm j}\,\nu}=-1$.

An interesting subset of these models is that in which only the dark matter is coupled to the dark energy, with the baryonic sector uncoupled. This corresponds to setting $\alpha_b(\phi)=0$ and $\alpha_c(\phi) = \alpha(\phi)$. Defining $\rho_{\rm j} =
e^{3 \alpha_{\rm j}} {\bar \rho}_{\rm j}$ gives
\bea
\mpl^2 G_{\mu\nu} &=& \nabla_\mu \phi \nabla_\nu \phi - \frac{1}{2} g_{\mu\nu}
(\nabla \phi)^2 - V(\phi) g_{\mu\nu}\nonumber
\\ && + e^{\alpha(\phi)}\rho_c u_{c\mu} u_{c\nu} \ ,
\label{ee}
\eea
and $
\nabla_\mu \nabla^\mu \phi - V^{\rm eff}_{,\phi}(\phi) = 0,
$
where we have defined an effective potential by
$
V_{\rm eff}(\phi) = V(\phi) + e^{\alpha(\phi)}\rho_c \ .
$
The fluid obeys $\nabla_\mu ( \rho_c u_c^\mu) =0$,
and $u_c^\nu \nabla_\nu u_c^\mu  = - (g^{\mu\nu} + u_c^\mu u_c^\nu) \nabla_\nu \alpha$.


There are a number of interesting effects in coupled models, including a surprising problem - the adiabatic instability~\cite{Bean:2007ny,Bean:2007nx,Valiviita:2008iv,He:2008si,Gavela:2009cy}. In the regime in which the scalar field adiabatically tracks the minimum of the effective potential, a careful study of linear perturbation theory implies that, within a certain regime of parameters, exponential growth can occur. 

\begin{figure}[t!]
\centering{
\includegraphics[width=5in]{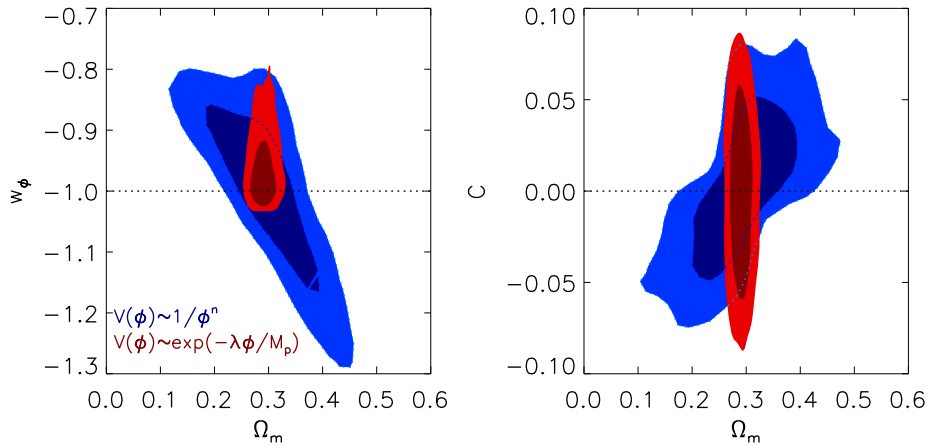}
\caption{Joint 68\% (dark shaded) and 95\% (light shaded) constraints   using combined WMAP CMB, SDSS matter power  spectrum,  SDSS and 2dFGRS baryon acoustic oscillation, `union' type
  1a supernovae datasets and HST $H_0$ prior for
  the power law potential (blue) and exponential potential (red)  for the  the fractional matter density, $\Omega_{\rm M}$ and the effective scalar equation of  state, $w_{\phi}$, (left panel), and the coupling, $C$ (right panel). From~\cite{Bean:2008ac}.
  \label{fig3}}}
\end{figure}

In general, one can constrain such coupled models using a combination of cosmological datasets. In~\cite{Bean:2008ac} the datasets used
included the measurements of the CMB temperature and polarization
power spectrum from the WMAP 5-year data release
\cite{Nolta:2008ih,Dunkley:2008ie},  the `union' set of supernovae
compiled by the Supernovae Cosmology Project  \cite{Kowalski:2008ez}, the matter power spectrum of Luminous
Red Galaxies (LRG) as measured by the Sloan Digital Sky Survey (SDSS)
survey \cite{Tegmark:2006az, Percival:2006gt} and a Gaussian prior on the Hubble constant today, $H_{0}=72\pm 8$, using the Hubble Space Telescope (HST) measurements~\cite{Freedman:2000cf}. The resulting constraint plots are shown in figure~(\ref{fig3}), expressed in terms of the dimensionless coupling variable $C$, defined by
\be
C(\phi)\equiv-\frac{M_p\alpha_{,\phi}}{\beta} 
\label{Cdef}
\ee
with $\beta \equiv \sqrt{2/3}$.


\section{Modified Gravity}
\label{sec:MGR}

As we have seen in the previous sections, within GR, a period of accelerated expansion can only be due to a dominant component of the energy density with a negative equation of state, the most common example being a static, or slowly rolling, scalar field (commonly referred to as {\it dark energy} or {\it quintessence} models). There is, of course, a natural alternative to be considered, namely the possibility that the dynamical rules determining how the geometry of spacetime responds to regular matter sources may differ from those given by GR. In other words, we may consider modifying GR in the low curvature regime, such as to admit self-accelerating solutions in the presence of negligible matter~\cite{Deffayet:2001pu,Dvali:2000hr,Freese:2002sq,Carroll:2006jn,Capozziello:2003tk,Carroll:2003wy,Deffayet:2000uy,Dvali:2003rk,Vollick:2003aw,Flanagan:2003rb,Flanagan:2003iw,Vollick:2003ic,Soussa:2003re,Nojiri:2003ni,ArkaniHamed:2003uy,Gabadadze:2003ck,Moffat:2004nw,Carroll:2004de,Clifton:2004st,Easson:2005ax,delaCruzDombriz:2006fj}.

While General Relativity has been tested to high precision in laboratory, in the solar system and astrophysically, for example with binary pulsars~\cite{Will:2005va}, there are far fewer tests on cosmological scales. It is therefore natural to explore whether  the laws of gravity could be modified on the larger scales and whether such modifications could be responsible for the late time acceleration. Obviously though, just as with dark energy approaches, models of modified gravity need to explain why, or assume a reason that the vacuum energy would be zero. With respect to the cosmological constant approach however, dark energy and modified gravity models have the advantage of adding dynamics to the theory, allowing for the possibility of alleviating the coincidence problem and increasing the experimental testability of the models.

When modifying gravity, one needs to ensure that GR is recovered at early times and on short scales, in regimes in which it has been well tested. Moreover, a viable model must agree with the precision measurements of the expansion history of the universe, of the CMB and of the growth of structure. As we shall see though, on theoretical grounds there are obstacles to constructing
infrared modifications of GR that will escape detection on smaller scales.
In a weak-field expansion around flat spacetime,
GR describes the propagation of massless spin-2 gravitons coupled to the
energy-momentum tensor $T_{\mu\nu}$, including that of the gravitons
themselves.  But such a theory is essentially unique; it has long been
known that we can start with a field theory describing a
transverse-traceless symmetric two-index tensor propagating in Minkowski
space, couple it to the energy-momentum tensor of itself and other
fields, and demonstrate iteratively that the background metric
disappears, leaving instead a curved metric obeying Einstein's equation.

It is therefore generally believed that infrared modifications
of GR will necessarily involve the introduction of new degrees of
freedom. This can be a good thing, in that the dynamics of these new degrees of freedom may be precisely what one needs to drive the accelerated expansion of the universe. However there are often problems associated with the extra dynamics and the extra forces mediated by these degrees of freedom. The problems may be of several different kinds. First, there is the possibility that, along with the desired deviations from GR on cosmological scales, one may also find deviations on smaller scales; e.g. in the solar system, where GR is well tested. Second is the possibility that the newly-activated degrees of freedom may be badly behaved in one way or another; either having the wrong sign kinetic terms ({\it ghosts}), and hence being unstable, or leading to superluminal propagation. More generally, theories that contain higher than first order derivatives in the Lagrangian risk being plagued by the instability described by the Ostrogradski theorem~\cite{Ostro,Woodard:2006nt}, i.e. the Hamiltonian of the system may be unbounded from below.  

The previous arguments put surprisingly restrictive constraints on models of modified gravity, and this has motivated some authors to explore models that modify GR without introducing new degrees of freedom~\cite{Carroll:2006jn,Afshordi:2007yx,Cooney:2008wk}.

In this section we will give an overview of some of the main candidate models of modified gravity, focusing on the general challenges that such theories face and the peculiar signatures to which they can lead in cosmology.

There are several ways in which one can imagine of modifying the laws of gravity on the largest scales. A first approach could perhaps focus only on the phenomenology of cosmology, in which case one postulates specific modifications of the equations describing the dynamics of the universe such as to achieve the acceleration. An example is the Cardassian model~\cite{Freese:2002sq}, in which the Friedmann equation is modified such that the dependence of the Hubble parameter on matter sources is different than the standard one. Although models such as this can yield cosmic acceleration, it is desirable to begin with a theory described by a generally covariant action, from which we may derive not only a modified Friedmann equation, but also the solution describing any other physical setting we may wish to know about. 

One broad class of modified gravity models that have been studied in detail contains those theories which involve covariant modifications of the $4$-dimensional action by adding higher order curvature invariants to the Einstein-Hilbert action. The simplest examples in this class are the so called $f(R)$ theories. It is instructive to go over some of the details of $f(R)$ models since they are representative of generalized modified theories of gravity. We will do so in the following subsection, while more details about these models can be found in the recent review by Faraoni {\it et al.}~\cite{Sotiriou:2008rp}. 

Finally, modifications of gravity can originate from extra dimensional scenarios, in which all matter fields are confined to a $4$-dimensional brane embedded in a higher dimensional bulk. The most prominent example is the Dvali-Gabadadze-Porrati (DGP) model~\cite{Dvali:2000hr}, in which the $4$D brane is embedded in a $5$D Minkowksi bulk of infinite volume, into which gravity may leak, weakening its strength. We dedicate the last subsection to this class of models and to some recent proposals that elaborate on them, e.g. {\it Degravitation}~\cite{Dvali:2007kt} and {\it Cascading Gravity}~\cite{deRham:2007rw,deRham:2007xp}. 

\subsection{$f(R)$ theories}
\label{fofrdescription}
It is a well known fact that the Einstein-Hilbert action is non renormalizable, therefore standard GR cannot be 
properly quantized. However renormalizability can be cured by adding to the action higher order terms in curvature invariants~\cite{Utiyama:1962sn, Vilkovisky:1992pb,Stelle:1976gc,Stelle:1977ry}. Moreover, higher order terms are generally expected in low energy effective actions derived from string theory or Grand Unification Theories (GUT). In 1979 Starobinsky showed that one can achieve a de Sitter phase in the early universe by adding to the Ricci scalar $R$ in the Einstein action a function $f(R)\propto R^2$~\cite{Starobinsky:1979ty,Starobinsky:1980te}. More recently, $f(R)$ models have been revisited in the context of cosmic acceleration, starting with~\cite{Capozziello:2003tk,Carroll:2003wy} and followed by many others. The underlying idea is to add functions of the Ricci scalar which become important at late times, for small values of the curvature.

Let us focus on the following action
\be
S=\frac{1}{2\kappa^{2}}\int d^4 x\sqrt{-g}\, \left[R+f(R)\right] + \int d^4 x\sqrt{-g}\, {\cal L}_{\rm m}[\chi_i,g_{\mu\nu}] \ ,
\label{jordanaction}
\ee
where $f(R)$ is a general function of the Ricci scalar, $R$. The matter Lagrangian, ${\cal L}_{\rm m}$, is minimally coupled and, therefore, the matter fields, $\chi_i$, fall along geodesics of the metric $g_{\mu\nu}$. The field equations obtained from varying the action~(\ref{jordanaction}) with respect to $g_{\mu\nu}$ are
\be \label{jordaneom}
\left(1+f_R \right)R_{\mu\nu} - \frac{1}{2} g_{\mu\nu} \left(R+f\right) + \left(g_{\mu\nu}\Box 
-\nabla_\mu\nabla_\nu\right) f_R =\kappa^{2}T_{\mu\nu} \ ,
\ee
where we have defined $f_R\equiv \partial f/\partial R$. We take the energy-momentum tensor to be that of a perfect fluid~(\ref{perfectfluid}). The metric $g_{\mu\nu}$ is minimally coupled to matter, hence the stress tensor and its conservation law will be the same as in standard GR. 

The $f(R)$ term in the gravitational action leads to extra terms in the Einstein equations. The constraint equation of GR becomes a third order equation, and the evolution equations now become fourth order differential equations (in contrast to the second order equations of standard GR). In particular, when taking the metric to be of the flat FRW form~(\ref{FRWmetric}), the Friedmann equation becomes
\ba
H^2+\frac{f}{6}-\f{\ddot{a}}{a}f_{R} +H\dot{f}_{R} &=&  \f{\kappa^{2}}{3}\rho
\label{jordanfriedmann}
\ea
and the acceleration equation is
\be\label{accelerationfofr}
\f{\ddot{a}}{a}-f_RH^2+\f{f}{6}+\f{\ddot{f}_R}{2}=-\f{\kappa^2}{6}(\rho+3P)\,.
\ee

Cosmic acceleration in equations~(\ref{jordanfriedmann}) and (\ref{accelerationfofr}) can arise through the additional terms present for $f(R)\ne 0$. One can interpret  these extra terms as the contribution from an {\it effective fluid} with an equation of state
\be\label{eos_eff}
w_{{\rm eff}}=-\f{1}{3}-\f{2}{3}\f{\l[H^2f_R-\f{f}{6}-H\dot{f}_R-\f{1}{2}\ddot{f}_R\r]}{\l[-H^2f_R-\f{f}{6}-H\dot{f}_R+\f{1}{6}f_RR\r]} \ .
\ee
It is important to notice that equation~(\ref{eos_eff}) can be used as a differential equation for the function $f(R)$. Indeed, given an expansion history, with a specified matter content, one can determine $w_{\rm{eff}}$ and then solve eq.~(\ref{eos_eff}) for $f(R)$. Since the resulting equation for $f(R)$ is of second order, one has a family of $f(R)$ models which reproduce the desired expansion history. In this sense, there is a degeneracy among models at the level of background cosmology, and, as we will discuss shortly, the growth of perturbations will play a key role in breaking this degeneracy. 

There exists a complementary, and sometimes conceptually simpler, way in which to approach $f(R)$ theories. It is possible to conformally transform the metric and bring the gravitational part of the action to the usual Einstein-Hilbert form of standard GR.  The price one pays for this simplification is a non-minimal coupling between matter fields and the transformed metric~\cite{Amendola:1999er,Bean:2000zm,Barrow:1988xh}, as well as the appearance of a new scalar degree of freedom playing the role of DE and evolving under a potential determined by the original form of the function $f(R)$ in~(\ref{jordanaction}). When the transformation between the two frames is well-defined, the classical results obtained using either description should be the same. The frame in which matter particles fall along geodesics of the metric, but in which the gravitational part of the action is modified, is referred to as the \emph{Jordan frame}. The frame in which the gravitational part of the action is the same as in standard GR, while the matter may be non-minimally coupled to gravity, is called the \emph{Einstein frame}.  

The emergence of an additional scalar degree of freedom can also be seen directly in the Jordan frame. The role of the scalar field is played by $f_R$, dubbed {\it the scalaron} in~\cite{Starobinsky:1980te}, and obeys an equation of motion given by the trace of equation~(\ref{jordaneom}), which can be written as
\be
\Box{f}_R={1\over 3}\left(R+2f-Rf_R \right)-{\kappa^2 \over 3}(\rho-3P) 
\equiv {\partial V_{\rm{eff}} \over \partial f_R} \ ,
\ee
which is a second order equation for $f_R$, with a canonical kinetic term and an effective  potential $V_{\rm{eff}}(f_R)$. Any $f(R)$ theory designed to achieve cosmic acceleration must satisfy $|f\ll R|$ and $|f_R|\ll 1$ at high curvatures to be consistent with our knowledge of the high redshift universe. In this limit, the extremum of the effective potential lies at the GR value $R=\kappa^2(\rho-3P)$. Whether this extremum is a minimum or a maximum is determined by the second derivative of $V_{\rm{eff}}$ at the extremum, which is also the squared mass of the scalaron
\be
m^2_{f_R} \equiv {\partial^2 V_{\rm{eff}} \over \partial f_R^2} = 
{1\over 3}\left[{1+f_R \over f_{RR}}-R \right] \ .
\ee

At high curvatures, when $|Rf_{RR}|\ll 1$ and $f_R\rightarrow 0$, 
\be
m^2_{f_R} \approx {1+f_R \over 3f_{RR}} \approx {1 \over 3f_{RR}}\ .
\ee
It then follows that in order for the scalaron not to be tachyonic, (i.e. to have a positive squared mass), one must require $f_{RR}>0$. Classically, $f_{RR}>0$ is required to keep the evolution in the high curvature regime stable against small perturbations \cite{Dolgov:2003px,Sawicki:2007tf}. 

The scalaron mediates an attractive ``fifth force'', which has a range determined by the Compton wavelength
\be
\label{lambda_C}
\lambda_c \equiv {2\pi \over m_{f_R}} \ .
\ee
While $\lambda_c$ is large at current cosmological densities, terrestrial, solar and galactic tests are not necessarily violated because the scalaron acquires a larger mass in regions of high matter density. This is essentially the Chameleon mechanism of~\cite{Khoury:2003aq,Khoury:2003rn}.

As we mentioned above, alternatively one may map the theory to the Einstein frame; following the approaches of~\cite{Chiba:2003ir,Barrow:1988xh,Magnano:1993bd}, one can recast the gravitational action~(\ref{jordanaction}) into a dynamically equivalent form by applying the conformal transformation
\be
{\tilde g}_{\mu\nu} = e^{2 \omega(x^{\alpha})} g_{\mu\nu} \ ,
\label{conftrans}
\ee
such that the function $\omega(x^{\alpha})$ satisfies
\be
e^{-2 \omega}(1+f_{R}) = 1 \ .
\label{constraint}
\ee
With this choice of $\omega$ the action~(\ref{jordanaction}) transforms into an action with the usual Einstein-Hilbert form for gravity. If we now define the scalar field  $\phi \equiv 2\omega / \beta\kappa$, where $\beta \equiv\sqrt{2/3}$, the resulting action becomes 
\ba
{\tilde S}=\frac{1}{2\kappa^{2}}\int d^4 x\sqrt{-{\tilde g}}\, {\tilde R} 
&+& \int d^4 x\sqrt{-{\tilde g}}\, 
\left[-\frac{1}{2}{\tilde g}^{\mu\nu}(\tilde{\nabla}_{\mu}\phi)\tilde{\nabla}_{\nu}\phi -V(\phi)\right] \nonumber\\
&+&\int d^4 x\sqrt{-{\tilde g}}\, e^{-2\beta\kappa\phi} {\cal L}_{\rm m}[\chi_i,e^{-\beta\kappa\phi}{\tilde g}_{\mu\nu}]\ ,
\label{einsteinaction}
\ea
where the potential $V(\phi)$ is determined entirely by the original form~(\ref{jordanaction}) of the action and is given by
\be
V(\phi)=\frac{1}{2\kappa^{2}}\frac{R f_{R} - f}{(1+f_{R})^{2}} \ ,
\label{einsteinpotential}
\ee
and where a tilde is used to indicate Einstein's frame quantities.

The field equations obtained by varying the action with respect to the metric ${\tilde g}_{\mu\nu}$ are
\ba
{\tilde G}_{\mu\nu}& =& 8 \pi G \tilde{T}_{\mu \nu} + \frac{1}{2} \tilde{\nabla}_{\mu} \phi \tilde{\nabla}_{\nu} \phi + \frac{1}{2}(\tilde{g}^{\alpha \gamma} \tilde{\nabla}_{\alpha} \phi \tilde{\nabla}_{\gamma} \phi ) \tilde{g}_{\mu\nu} - V(\phi) \tilde{g}_{\mu\nu} \label{einsteineom1}\\
&=&8 \pi G\l( \tilde{T}_{\mu \nu}+T_{\mu\nu}^{\phi}\r) \ ,
\label{einsteineom2}
\ea
where $\tilde{T}_{\mu \nu}$ is the energy-momentum tensor of matter fields in the Einstein frame. In the remaining terms on the right hand side of~(\ref{einsteineom1}) we recognize the energy-momentum tensor $T_{\mu\nu}^{\phi}$ of the scalar field $\phi$. Although quantities in the Einstein frame are more familiar than those in the Jordan frame, there are some crucial distinctions.  Most notably, in this frame test matter particles do not freely fall along geodesics of the metric ${\tilde g}_{\mu\nu}$, since the scalar field is also coupled to matter, as can be seen from the conservation equations for the scalar field and for the perfect fluid matter, which read
\ba
\ddot{\phi} + 3\tilde{H}\dot{\phi} + V_{,\phi} = \f{1}{2}\kappa\beta (\tilde{\rho} - 3 \tilde{P} ) \ ,
\label{scalareom}
\\
\dot{\tilde{\rho}} +3\tilde{H}(\tilde{\rho} + \tilde{P}) = - \f{1}{2}\kappa\beta \dot{\phi}( \tilde\rho-3\tilde P) \ ,
\ea
where  $V_{\phi}\equiv dV/d\phi$.

From this mapping it is clear that $f(R)$ theories are a specific example of the broader class of scalar-tensor theories that we discussed in the final subsection of the previous section. 

Returning to the Jordan frame formulation, there is a series of conditions that must be imposed on the function $f(R)$ in order to avoid instabilities and ensure agreement with local tests of gravity. These conditions can be summarized as follows
\begin{enumerate}
\item \label{cond:1} $f_{RR}>0$ for $R\gg f_{RR}$. Classically, this follows from requiring the existence of a stable high-curvature regime, such as the matter dominated universe~\cite{Dolgov:2003px}. 

\item \label{cond:2} $1+f_R>0$ for all finite $R$. The most direct interpretation of this condition is that the effective Newton constant, $G_{\rm{eff}}=G/(1+f_R)$, is not allowed to change sign. This prevents the graviton from becoming ghost-like~\cite{Nunez:2004ji}. 

\item \label{cond:3} $f_R<0$. Given tight constraints from Big Bang nucleosynthesis (BBN) and the CMB, we must require GR to be recovered at early times, i.e. $f(R)/R\rightarrow 0$ and $f_R \rightarrow 0$ as $R\rightarrow \infty$. Together with $f_{RR}>0$, this implies that $f_R$ must be a negative, monotonically increasing function of $R$ that asymptotes to zero from below.

\item \label{cond:chameleon} $f_R$ must be small at recent epochs. This is not required if the only aim is to build a model of cosmic acceleration, without trying to satisfy the solar and galactic scale constraints. This condition ensures a small difference between the value of the scalaron field in the high density galactic center and the value in the outskirts of the galaxy~\cite{Hu:2007nk}. As shown in~\cite{Khoury:2003rn}, the difference between these values is effectively the potential difference that sources the attractive ``fifth force'' acting on objects in the vicinity of the galaxy. The analysis of~\cite{Hu:2007nk} suggests that the value of $|f_R|$ today should not exceed $10^{-6}$. However, as stressed in~\cite{Hu:2007nk}, this bound assumes that in $f(R)$ models galaxy formation proceeds similarly to that in GR. Hence, while it is certain that $|f_R|$ must be small, any specific bound on its value today will be unreliable until galaxy formation in $f(R)$ is studied in N-body numerical simulations - work that has only recently begun~\cite{Oyaizu:2008tb,Schmidt:2008tn}.
\end{enumerate}

Finally, it was recently shown~\cite{Frolov:2008uf,Kobayashi:2008tq} that even models of $f(R)$ that satisfy all of the above conditions might still face problems in the strong gravity regime. These problems are due to a curvature singularity $R\rightarrow 0$ appearing at the non-linear level, at a finite field value and energy, such that the scalar degree of freedom can easily access it in the presence of matter. As a consequence, $f(R)$ theories appear to be incompatible with the existence of static dense compact objects like neutron stars. The problem seems to require an ultraviolet completion of $f(R)$ theories, and indeed it was recently shown in~\cite{Dev:2008rx,Kobayashi:2008wc} that the addition of higher order curvature corrections might solve the curvature singularity problem, although an unpleasant degree of fine-tuning would be required. 

\subsection{Generalized theories of gravity}
The $f(R)$ theories studied in the previous subsection are the simplest example of generalized theories of gravity. Still, in their simplicity, they retain some of the features of more general higher order theories of gravity. Studying the cosmology of these models, we have learned how the introduction of higher order terms in curvature invariants can lead to instabilities and/or conflicts with local tests of gravity because of extra degrees of freedom that come into play. These problems are characteristic of generalized theories of gravity, and can be more severe in models that involve functions of curvature invariants other than the Ricci scalar.

There are, of course, any number of terms that one could consider, but for simplicity, we might consider those invariants of lowest mass dimension that are also parity-conserving
\begin{eqnarray}
P &\equiv & R_{\mu\nu}\,R^{\mu\nu} \ \nonumber \\
Q &\equiv & R_{\alpha\beta\gamma\delta}\,R^{\alpha\beta\gamma\delta} \ .
\end{eqnarray}

and actions of the form~\cite{Carroll:2004de}
\begin{equation}
S=\int d^4x \sqrt{-g}\,[R+f(R,P,Q)] +\int d^4 x\, \sqrt{-g}\,
{\cal L}_M \ ,
\label{genaction}
\end{equation}
where $f(R,P,Q)$ is a general function describing the deviations from general relativity.

It is convenient to define
\begin{equation}
f_R\equiv\frac{\partial f}{\partial R}\ , \qquad
f_P\equiv\frac{\partial f}{\partial P}\ , \qquad 
f_Q\equiv\frac{\partial f}{\partial Q} \ ,
\end{equation}
in terms of which the equations of motion are
\begin{eqnarray}
R_{\mu\nu} &-& \frac{1}{2}\,g_{\mu\nu}\,R-\frac{1}{2}\,g_{\mu\nu}\,f \nonumber \\
&+& f_R\,R_{\mu\nu}+2f_P\,R^\alpha{}_\mu\,R_{\alpha\nu}
+2f_Q\,R_{\alpha\beta\gamma\mu}\,R^{\alpha\beta\gamma}{}_\nu\nonumber\\
&+& g_{\mu\nu}\,\Box f_R -\nabla_\mu\nabla_\nu f_R
-2\nabla_\alpha\nabla_\beta[f_P\,R^\alpha{}_{(\mu}{}\delta^\beta{}_{\nu)}]
+\Box(f_P\,R_{\mu\nu})\nonumber\\
&+& g_{\mu\nu}\,\nabla_\alpha\nabla_\beta(f_P\,R^{\alpha\beta})
-4\nabla_\alpha\nabla_\beta[f_Q\,R^\alpha{}_{(\mu\nu)}{}^\beta]
=8\pi G\,T_{\mu\nu}\ .\label{equaz}
\end{eqnarray}

It is straightforward to show that actions of the form~(\ref{genaction}) generically admit a maximally-symmetric solution: $R=$ a non-zero constant. However, an equally generic feature of such models is that this de Sitter solution is unstable (corresponding to one of the new degrees of freedom attaining a local potential maximum), and in many cases there exists another accelerating solution, analogous to those that we found for $f(R)$ models.

What about constraints on these models? It has been shown~\cite{Navarro:2005gh} that solar system constraints, of the type we have described for $f(R)$ models, can be evaded by large classes of these more general models. Roughly speaking, this is because the Schwarzschild solution, which governs the solar system, has vanishing $R$ and $P$, but non-vanishing $Q$.

Far more serious though is the issue of ghosts and superluminal propagation. It has been shown~\cite{Chiba:2005nz,Navarro:2005da} that a necessary but not sufficient condition that the action be ghost-free is that $b=-4c$, so that there are no fourth derivatives in the linearized field equations. What remained was the possibility that the second derivatives might have the wrong signs, and also might allow superluminal propagation at some time in a particular cosmological background. It has recently been shown that around a FRW background with matter, the theories are ghost-free, but contain superluminally propagating scalar or tensor modes over a wide range of parameter space~\cite{DeFelice:2006pg,Calcagni:2006ye}. However, more generally it has been argued~\cite{Woodard:2006nt}, that only modifications of the action that are functions of the Ricci scalar and the Gauss-Bonnet term can avoid the instability of the Ostrogradski's theorem. Thus, such general actions do not seem to be promising candidates for modified gravity theories explaining cosmic acceleration.

\subsection{Modifying gravity without extra degrees of freedom}

The challenges that higher order theories of gravity theories face are related to the dynamics of, and forces mediated by, the extra degree(s) of freedom introduced by the modifications. It is therefore interesting to investigate whether it is possible to achieve cosmic acceleration with models for which the constraint equations of gravity are modified while the graviton is still the only propagating degree of freedom~\cite{Carroll:2006jn,Afshordi:2007yx}. Since such models are a little outside the scope of this review, we will not comment on them further, other than to point out that linear perturbation theory seems to show a possible conflict with observations of the formation of structure~\cite{Carroll:2006jn}.

\subsection{The DGP model and its derivatives}
So far we have considered models of modified gravity based on a covariant four-dimensional action. In this subsection we shall review some of the proposals in the context of extra dimensions. We focus on brane-world models that modify gravity at low energies; i.e. models that can be used to achieve cosmic acceleration. 

The most prominent proposal in this context is the well known Dvali-Gabadadze-Porrati (DGP)~\cite{Dvali:2000hr} model, the cosmology of which was described in~\cite{Deffayet:2001pu}. In this model, the matter fields are constrained to lie on a $4D$ brane embedded in a flat $5$D Minkowski bulk of infinite volume, into which the graviton can leak. In the DGP model gravity appears four dimensional at short distances, while it is modified on distances large compared to a crossover scale $r_c$ through the evaporation of the graviton into the unseen fifth dimension. 

The $5$D action describing the model is 
\be\label{DGPaction}
S_5=-\f{1}{16\pi}M^3\int\,d^5x\,\sqrt{-g}\,R\,-\f{1}{16\pi}M_P^2\int\,d^4x\,\sqrt{-g^{(4)}}\,\l[R^{(4)}-\f{16\pi}{M_P^2}{\mathcal L}_{\rm{m}}\r]
\ee
where $g$, $M$ and R represent, respectively, the five dimensional metric, Planck mass and Ricci scalar, while $g^{(4)}$, $M_P$ and $R^{(4)}$ are their four dimensional counterparts. Finally,  ${\cal L}_m$ is the $4D$ matter Lagrangian density.  The crossover scale $r_c$ is given by the ratio of the observed $4$D Planck mass to the  $5$D one
\be
r_c=\frac{M_P^2}{2M^3} \ .
\ee 
This represents the distance over which metric fluctuations propagating on the brane dissipate into the bulk, weakening gravity. As a consequence, the $4$D effective gravitational potential will behave as
\be\label{DGPpotential}
\Phi\propto\left\{\begin{array}{cc}r^{-1},& \,\,\, r\ll r_c\\r^{-2},& \,\,\, r\gg r_c\end{array}\right. \ .
\ee

Interestingly, significant departures from standard gravity persist down to scales smaller than $r_c$; for a mass source of Schwarzschild radius $r_g$, modifications will be relevant down to scales equal to the Vainshtein radius $r_*\equiv (r_g r_c^2)^{1/3}$~\cite{Vainshtein:1972sx,Dvali:2006su}.

In order to address the phenomenon of cosmic acceleration it is necessary to choose $r_c\sim H_0^{-1}$. The Friedmann equation describing the expansion of a $4$D brane with spatial curvature $k$ reads
\be\label{DGPFriedmann}
H^2+\f{k}{a^2}-\epsilon\f{1}{r_c}\sqrt{H^2+\f{k}{a^2}}=\f{8\pi G}{3}\rho\,.
\ee
where $\epsilon=\pm1$ correspond to the two different branches of the model. The $\epsilon=+1$ choice corresponds to the self-accelerated branch, {\it sDGP}; this model naturally leads to the late time acceleration and has been recently tested against SNeIa, BAO and CMB data~\cite{Maartens:2006yt}. It was found that the flat sDGP ($k=0$) model, is a poor fit to the data and that the $\Lambda$CDM model gives a significantly better fit~\cite{Sahni:2002dx,Alam:2002dv,Alam:2005pb,Lazkoz:2006gp,Lazkoz:2007zz}. Moreover, sDGP is plagued by the presence of a ghost in the gravitational background sector, which is a serious problem.

The choice $\epsilon=-1$ corresponds to the normal branch, {\it nDGP}, which is obtained via a different choice of embedding of the brane into the bulk. It requires the addition of a cosmological constant to achieve cosmic acceleration, but it has the advantage of being ghost free. In this model, the cosmological constant is  screened by $5$D effects, and therefore it is the simplest example of {\it degravitation}, which we will mention shortly. However the screening effect is too small to solve the fine tuning problem of the cosmological constant. Another interesting feature of nDGP is that it displays an effective phantom behavior~\cite{Sahni:2002dx,Alam:2002dv,Alam:2005pb,Lazkoz:2006gp}. A comparison of the nDGP model with a combination of cosmological datasets~\cite{Sahni:2002dx,Alam:2002dv,Alam:2005pb,Lazkoz:2006gp,Lazkoz:2007zz}, found that closed models were a good fit to the data. 

In~\cite{Dvali:2007kt} it was pointed out that massive-graviton theories, such as DGP, could lead to {\it degravitation} of the vacuum energy,  offering a dynamical solution to the cosmological constant problem. Degravitation indeed allows for a large cosmological constant, but suppresses its backreaction by making gravity sufficiently weak on large scales. The degravitation of the vacuum energy is an intriguing solution to the fine tuning problem of the cosmological constant, and can be achieved if Newton's constant is promoted to a derivative operator~\cite{Dvali:2007kt}
\be\label{degravitation_Einstein_Eqs}
G^{-1}\,\l(L^2\Box\r)\,G_{\mn}=8\pi\, T_{\mn}\,.
\ee
The gravity described by the above Einstein equations will behave as a high-pass filter with characteristic scale $L$, such that $G\rightarrow  G^{(0)}$ for $L^2(-\Box)\rightarrow\infty$, and $G\rightarrow 0$ for $L^2(-\Box)\rightarrow 0$. Equation~(\ref{degravitation_Einstein_Eqs}) should be considered only as a phenomenological description of the degravtitaion model, the details of which can be found in~\cite{Dvali:2007kt}.

Although DGP is effectively a massive graviton theory, the degraviation phenomenon cannot be achieved in the context of standard DGP because the weakening of gravity is not sufficiently effective. A very promising way of realizing degravitation is through the so called {\it Cascading Gravity} model~\cite{deRham:2007rw,deRham:2007xp,Kaloper:2007ap,Kaloper:2007qh}. In this model, DGP is extended to higher dimensions, in which our visible $3$-brane is embedded within a succession of higher-dimensional branes, each with their own induced gravity terms, embedded in a flat $D$-dimensional bulk. The model is free of ghost-like instabilities (common to previous attempts at a higher-codimenion DGP) and the $4$D propagator is well behaved when calculated on the brane. Moreover, this model exhibits degravitation with a steep enough weakening of gravity -- the gravitational force law {\it cascades} from a $1/r^2$ ($4$D) behavior, to $1/r^3$, ($5$D) to $1/r^4$, ($6$D) behavior, etc. as one probes larger distances on the $3$-brane.

These models are works in progress, but are particularly interesting examples of modified gravity models in which both cosmic acceleration and the cosmological constant problem may be addressed.


\section{Distinguishing between models}
\label{sec:tests}
The $\Lambda$CDM model is currently the best fit to the available cosmological data, and passes all local tests of gravity~\cite{Will:2005va}. As we have already discussed,  it is however important to explore
 the whole space of explanations that fit the data comparably well and have testable features. In the previous sections we saw that, in general, imposing cosmological and local viability of models of modified gravity, as well as generalized DE, requires a degree of fine-tuning. After this tuning,  the predictions of these models for the expansion history of the universe are often identical to that of the $\Lambda$CDM model~\cite{Dvali:2007kt,Hu:2007nk,Appleby:2007vb,Pogosian:2007sw,Brax:2008hh,Capozziello:2008fn}. 

However, this degeneracy is typically broken at the level of cosmological structure formation; indeed, models of modified gravity that closely mimic the cosmological constant at the background level can nevertheless yield significantly different predictions for the growth of structure. In recent times it has become increasingly evident, that the large scale structure of the universe offers therefore a promising testing ground for GR. To this extent, it is important to explore to what degree one can use LSS to distinguish among all the candidate models of cosmic acceleration and, indeed, whether model-independent approaches are possible~\cite{Daly:2004gf,Daly:2003iy,Sahni:2006pa,Daly:2007dn}.

Scalar metric perturbations in the Newtonian gauge are described by two potentials, $\Psi(\vec{x},t)$ and $\Phi(\vec{x},t)$, defined via the perturbed FRW line element
\be\label{metric}
ds^2=-\l(1+2\Psi(\vec{x},t)\r)dt^2+a^2(t)\l(1-2\Phi(\vec{x},t)\r)d\vec{x}^2 \ .
\ee
Thus, these potentials correspond, respectively, to perturbations in the time-time and space-space components of the metric tensor, representing the strength of gravity and the spatial curvature. In the $\Lambda$CDM model, the potentials are equal during the epoch of structure formation, and their time dependence is set by the same scale-independent linear growth function that describes the growth of matter density perturbations. 
This, it turns out, is a very peculiar feature of the $\Lambda$CDM model, and of models of dark energy with negligible shear and clustering. However, it no longer holds in theories of modified gravity or models of coupled dark energy, where one can have scale-dependent growth patterns. In these theories, the two Newtonian potentials generically differ, and their dependence on matter perturbations can be different. 
As a consequence, modifications of gravity generally introduce a time- and scale-dependent slip between the Newtonian potentials, as well as a time- and scale-dependent effective Newton's constant describing the clustering of dark matter.  These modifications are expected to leave distinct imprints on the large scale structure of the universe, which may help to break the degeneracy that characterizes models of cosmic acceleration at the background level. This explains the recent growing interest in studying the potential of current and upcoming surveys to detect and constrain any departure from GR in the growth of structure~\cite{Caldwell:2007cw,Dore:2007jh,Bertschinger:2008zb,Daniel:2008et,Zhang:2007nk,Zhang:2008ba,Zhao:2008bn,Song:2008xd}. 

\subsection{Modified Growth of Structure}
It is useful to review  the evolution of linear scalar perturbations in general models of modified gravity  or coupled dark energy, as it will allow us to describe in detail some of the recent efforts to constrain departures from GR using the growth of structure.

We focus on the evolution of linear matter and metric perturbations in a general metric theory of gravity. Assuming that the background evolution is correctly described by the flat FRW metric, we consider scalar perturbations in Newtonian gauge, as defined in~(\ref{metric}). Representing all perturbed quantities in Fourier space, we use the standard notation for the energy momentum tensor of the matter fields, which to first order in the perturbations, assumes the form
\ba\label{en-mom_tensor}
&&{T^0}_0=-\rho(1+\delta),\nonumber\\
&&{T^0}_i=-(\rho+P)v_i,\\
&&{T^i}_j=(P+\delta P)\delta^i_j+\pi^i_j,
\ea
where $\delta\equiv \delta\rho/\rho$ is the density contrast, $v$ the velocity field, $\delta P$ the pressure perturbation and $\pi^i_j$ denotes the traceless component of the energy-momentum tensor. Finally, we define the anisotropic stress $\sigma$ via $(\rho+P)\sigma\equiv -(\hat{k}^i\hat{k}_j-\f{1}{3}\delta^i_j)\pi^i_j$.

In GR, the linearized Einstein equations provide two independent equations relating the metric potentials and matter perturbations, the {\it Poisson} and {\it anisotropy} equations, respectively:
\ba\label{Poisson}
&&k^2\Phi=-\f{a^2}{2M_P^2}\rho\Delta \ ,\\
\label{anisotropy}
&&k^2(\Phi-\Psi)=\f{3a^2}{2M_P^2}(\rho+P)\sigma \,,
\ea
where $\rho\Delta\equiv\rho\delta+3\f{aH}{k}(\rho+P)v$ is the comoving density perturbation.
In the $\Lambda$CDM and minimally coupled quintessence models, the anisotropic stress is negligible at times relevant for structure formation, and we have $\Psi=\Phi$.

In models of modified gravity, as well as in more exotic models of dark energy, the relation between the two Newtonian potentials, and between the potentials and matter perturbations, can be different~\cite{Zhang:2005vt,Bertschinger:2006aw,Song:2006ej}. The effect of these modifications can be encoded in two functions of time and space, a rescaling of the Newton's constant $\mu(a,k)$ and a {\it gravitational slip} $\gamma(a,k)$, defined via
\ba\label{parametrization-Poisson}
&&k^2\Psi\equiv-\f{a^2}{2M_P^2}\mu(a,k)\rho\Delta\\
\label{parametrization-anisotropy}
&&\f{\Phi}{\Psi}\equiv\gamma(a,k) \ ,
\ea
where $\mu(a,k)$ and $\gamma(a,k)$ are time- and scale-dependent functions. Note that $\mu$ is defined via the Poisson equation (\ref{parametrization-Poisson}) written in terms of $\Psi$, the perturbation to the time-time component of the metric. This choice is natural, as it is $\Psi$ that enters the evolution equation for CDM density perturbations on sub-horizon scales:
\be
\label{density_evol}
\ddot{\delta}+2H\dot{\delta}+\f{k^2}{a^2}\Psi=0 \ .
\ee
With the set of equations~(\ref{parametrization-Poisson}),~(\ref{parametrization-anisotropy}) and~(\ref{density_evol}), completed by an expression for $\mu$ and $\gamma$, one can evolve the linear perturbations forward in time and extract predictions for the growth of structure.
\subsubsection{An example}
\begin{figure}[t]
\begin{centering}
\includegraphics[width=140mm]{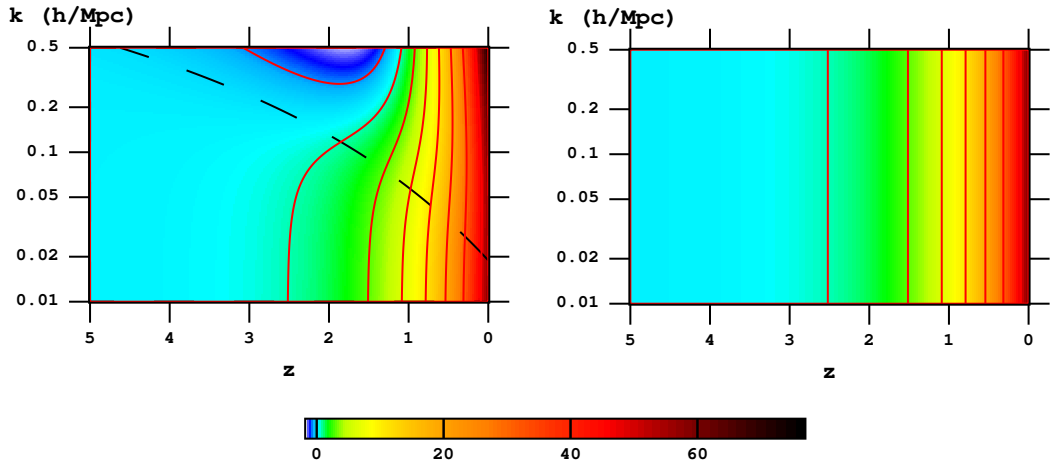}
\caption{The function probed by the cross-correlation of large scale structure with ISW, $\delta\cdot d(\Phi+\Psi)/dz$, as a function of scale $k$ and redshift $z$. The left panel corresponds to an $f(R)$ model with a $\Lambda$CDM expansion history and $f_R^0=-10^{-4}$ and shows a characteristic scale-dependent pattern. The right panel corresponds to $\Lambda$CDM. The dashed line crossing through the left panel corresponds to $\lambda=\lambda_c$~(\ref{lambda_C}), i.e. it  corresponds to the Compton wavelength of the scalaron  $\lambda_c$ (\ref{lambda_C}). For scales $\lambda<\lambda_c$, during matter domination, one can clearly notice the effect of the ``fifth force" which suppresses the cross correlation and can actually make the correlation negative. Therefore, a negative cross correlation signal at early redshifts (corresponding to matter era), is a signature of $f(R)$. The acceleration of the background will eventually contrast the ``fifth force" and lead to a positive cross-correlation.}
\label{ISW_f(R)}
\end{centering}
\end{figure}

It is instructive to look at the growth of structure in  a specific model of modified gravity. For this purpose, we will briefly review some results for the evolution of linear perturbations in $f(R)$ theories. The growth of structure in these models has been studied in~\cite{Pogosian:2007sw,Song:2006ej,Bean:2006up}, where it was found that the perturbations grow with a characteristic scale-dependent pattern. The Compton wavelength of the  scalaron (\ref{lambda_C}) introduces a scale which separates two regimes of sub-horizon gravitational dynamics during which gravity behaves differently. On scales $\lambda\gg\lambda_c$, the scalaron is massive and the ``fifth force'' it mediates  is exponentially suppressed; thus deviations from GR are negligible. However, on scales inside the Compton radius, the scalaron is light and deviations are significant. The relations between $\Phi$ and $\Psi$, and the relation between them and the matter density contrast, will be different below the Compton scale and that affects the growth rate of structure. Indeed, on sub-horizon scales ($k\gg aH$), as an effect of the modifications to the Einstein-Hilbert action, the Poisson and anisotropy equations read
\ba
\label{f(R)_Poisson}
\f{k^2}{a^2}\l(1+f_R\r)\Phi = -\f{a^2}{2M_P^2}\rho\Delta &+& \l[\f{1}{2}\f{k^2}{a^2}f_{RR}\delta R\r.\nonumber\\
&-&\l.\f{3}{2}\dot{f}_R\l(H\Psi+\dot{\Phi}\r)-\f{3}{2}\dot{H} f_{RR}\delta R\r]
\ea
\be
(1+f_R)\l(\Phi-\Psi\r) = f_{RR}\,\delta R \ ,
\label{f(R)_anisotropy}
\ee
where we have neglected any shear from matter fields. From the sub-horizon limit of equations~(\ref{f(R)_Poisson}),~(\ref{f(R)_anisotropy}), we can easily read off the resulting rescaling of the Newton's constant and the gravitational slip as
\ba\label{mu_f(R)}
\mu(a,k)&=&\f{a^2(1+f_R)+4f_{RR}k^2}{a^2(1+f_R)+3f_{RR}k^2}\simeq\f{\lambda^2+4\lambda_c^2}{\lambda^2+\lambda_c^2}\\
\label{gamma_f(R)}
\gamma(a,k)&=&\f{a^2(1+f_R)+2f_{RR}k^2}{a^2(1+f_R)+4f_{RR}k^2}\simeq\f{3\lambda^2+2\lambda_c^2}{3\lambda^2+4\lambda_c^2} \ ,
\ea
(where we have neglected a prefactor $(1+f_R)^{-1}$ multiplying the expression for $\mu$, since it is very close to unity in models that satisfy local tests of grvaity~\cite{Hu:2007nk}).
In particular, on scales below $\lambda_c$, the modifications introduce a slip between the Newtonian potentials, leading to the low-curvature regime $\Psi\simeq 2\Phi$.  The rate of growth of structure depends on the balance between the fifth force and the background acceleration. For modes that cross below $\lambda_c$ during matter domination, the effect of the modifications is maximized as the potentials grow in the absence of background acceleration. When, however, the background starts accelerating, the potentials begin to decay but at a lesser rate than in the $\Lambda$CDM model. Therefore, a characteristic signature of $f(R)$ theories would be a non-zero ISW effect during the matter era~\cite{Song:2006ej,Pogosian:2007sw}. The ISW constitutes just a small fraction of the overall CMB power on the larger scales, where cosmic variance dominates, and therefore it is very hard to extract. The way to proceed is to cross-correlate the CMB with galaxies, since one expects a correlation between the distribution of galaxies (i.e. DM potential wells) and the ISW signal~\cite{Giannantonio:2008zi,Crittenden:1995ak,Boughn:2003yz}. In $f(R)$ theories such a correlation would be negative, since the potentials would be growing. In Fig.~(\ref{ISW_f(R)}) we show such cross correlation for an $f(R)$ model, while some constraints using existing ISW data can be found in~\cite{Song:2007da}. One should note, however, that the ISW measurements are plagued by large statistical errors (because the ISW effect is only a part of the total CMB anisotropy) and cannot provide the percent level accuracy needed to test viable $f(R)$ models, (i.e. models that satisfy local tests of gravity and for which the ISW contribution is suppressed to the percent level). 
Weak lensing studies, on the other hand, are counted on to eventually provide highly accurate $3$D maps of the sum of the gravitational potentials, $\Phi+\Psi$.  

The features found in $f(R)$ models are common to the broader class of scalar-tensor theories, to which $f(R)$ belongs, as well as to models of coupled quintessence~\cite{Amendola:2003wa,Bean:2001ys}. These models can be described by a Chameleon-type action~(\ref{action0}), representative of coupled dark energy models in the Jordan frame and of scalar-tensor theories in the Einstein frame. Focusing, as before, on dark matter, indicating its coupling simply with $\alpha$ (i.e. dropping the index $i$), and studying 
scalar linear perturbations around a flat FRW metric, on sub-horizon scales we can easily find~\cite{Pogosian:2007sw,Song:2006ej} that the effective Newton's constant and the gravitational slip are
\ba\label{G_scalartensor}
&&\mu(a,k)=\f{1+\l(1+2M_P^2{\alpha^2_{,\phi}}\r)\f{k^2}{a^2m^2}}{1+\f{k^2}{a^2m^2}}\\
\label{gamma_scalartensor}
&&\gamma(a,k)=\f{1+\l(1-2M_P^2{\alpha^2_{,\phi}}\r)\f{k^2}{a^2m^2}}
{1+\l(1+2M_P^2{\alpha^2_{,\phi}}\r)\f{k^2}{a^2m^2}} \ ,
\ea
which are exactly of the form~(\ref{mu_f(R)})-(\ref{gamma_f(R)}). (In~(\ref{G_scalartensor}) we have neglected e prefactor $e^{-2\alpha(\phi)}\simeq 1$). Here $m$ is the mass of the scalar field and $\alpha_{,\phi}\equiv \partial\alpha/\partial\phi$.

To summarize, in models of modified gravity and coupled dark energy the growth pattern is in general scale-dependent  and the time evolution of the Newtonian potentials is different than that in $\Lambda$CDM. The introduction of an effective Newton's constant and a slip between the potentials is a general feature of the modifications, independent of the background expansion. Therefore, even for models that at the background level 
 closely mimic the cosmological constant, the predictions for the growth of structure can be significantly different. The degeneracy characterizing models of cosmic acceleration at the background level can henceforth be broken at the level of large scale structure. Characteristic imprints of such modifications will be a different clustering rate for DM, a modified ISW signal as well as a different Weak Lensing shear (WL). Moreover, since the potentials $\Phi$ and $\Psi$ will in general differ, one expects that the function measured by ISW and WL, i.e. $(\Phi+\Psi)$, will be different than the one to which   peculiar velocities respond, i.e. $\Psi$. All these features allow potentially powerful ways to distinguish between models of cosmic acceleration, which we review in the next subsection.

Before concluding, let us notice that, as we saw, models of generalized dark energy can induce a pattern of growth analogous to the one found in models of modified gravity. Therefore, the degeneracy between dynamical dark energy and modified gravity models appears to persist at the level of LSS. Indeed, it seems always possible to introduce a generic dark fluid with properties such as to mimic the modifications of gravity. This has been discussed recently by several groups~\cite{Kunz:2006ca,Hu:2007pj,Jain:2007yk} and the authors of~\cite{Jain:2007yk} have noted that the breaking of such degeneracy would require the independent measurements of three observables, (e.g. $\Psi$, $\Phi+\Psi$ and $\delta$).

\subsection{Constraining departures from  standard GR}
From the preceding discussion it is clear that LSS offers a potentially powerful way of constraining models of cosmic acceleration. The task, however, is not trivial, and there are different ways in which one can proceed. 
A first approach could perhaps consist of assuming a functional form for the slip between the potentials and the effective Newton's constant, motivated by a certain class of theories. Some representative parametrizations have been introduced in~\cite{Bertschinger:2008zb,Hu:2007pj,Hu:2008zd,Amin:2007wi}. From the previous examples we can easily infer that a quite generic parametrization would be
\ba\label{par}
&&\mu(a,k)=\f{1+\beta_1\lambda_1^2\,k^2a^s}{1+\lambda_1^2\,k^2a^s}\\
&&\gamma(a,k)=\f{1+\beta_2\lambda_2^2\,k^2a^s}{1+\lambda_2^2\,k^2a^s} \ ,
\ea
where the parameters $\lambda^2_i$ have dimensions of length squared, while the $\beta_i$ represent dimensionless couplings.  Finally, from a scalar-tensor point of view, the parameter $s$ encodes the time-dependence of the scalar field mass. The expressions in~(\ref{par}) coincide with the scale-dependent parametrization introduced in~\cite{Bertschinger:2008zb}. In~\cite{Zhao:2008bn}, the authors performed a Fisher matrix analysis to forecast the constraints on the parameters $\{\lambda_i,\beta_i,s\}$ of~(\ref{par})  from a combination of power spectra from galaxies, Weak Lensing and the Integrated Sachs-Wolfe effect on the CMB. Such an analysis reveals the extent to which one can constrain these theories and also allows one to reconstruct the shape of the slip and effective Newton's constant based on the chosen form. In~\cite{Zhao:2008bn}, it was found that these data, even at the linear level, are quite powerful in constraining the modified growth parameters. The results, of course, depend on the choice of the parametrization. However, they are good indicators of the power of current and upcoming surveys to constrain departures from GR. 

An alternative approach, which can work for certain estimators of the slip, is a direct reconstruction from data. In~\cite{Zhang:2007nk,Zhang:2008ba}, it was proposed to consider the ratio of the peculiar velocity-galaxy correlation with the Weak Lensing-galaxy correlation. Specifically, the authors propose the following estimator for the gravitational slip
\be\label{slip_Dodelson}
\hat{\gamma}=\f{P_{\nabla^2(\Psi-\Phi)g}}{P_{\nabla^2\Psi g}}-1\,,
\ee
where $P_{\nabla^2(\Psi-\Phi)g}$ and $P_{\nabla^2\psi g}$ are the cross-power spectrum between the galaxy number overdensity and, respectively, weak lensing and the potential $\Psi$ (which is related to peculiar velocities via the continuity equation).

In such a ratio, the dependence on the galaxy bias cancels out. Such a ratio, if appropriately constructed, would directly probe any difference between $\Phi$ and $\Psi$. This is a more direct and model-independent way of testing GR with the growth of structure; however its power will depend on how well future experiments will be able to measure  peculiar velocities. From the analysis in~\cite{Zhang:2007nk,Zhang:2008ba}, it appears however that the ratio measured by future surveys might have strong discriminating power for some dark universe scenarios.

Finally, another, non-parametric approach, consists of performing a \emph{Principal Component Analysis} (PCA) to determine the eigenmodes of the slip and the Newton constant, that can be constrained by data, in the same way as has been done for the dark energy equation of state~\cite{Huterer:2002hy,Crittenden:2005wj,dePutter:2007kf}. This method
allows one to compare different experiments and their combinations, according to the relative gain in information about the functions. PCA can also point to the ``sweet spots'' in redshift and scale where data is most sensitive to variations in the slip and Newton's constant, which can be a useful guide for designing future observing strategies.  The PCA method does not allow one to reconstruct the shape of the functions from data. However, one can still reproduce the errors on parameters of any parametrization from the eigenvectors and eigenvalues found using PCA~\cite{Crittenden:2005wj}. Thus, it is a promising method, but computationally challenging.
\subsubsection{The observables}
\label{tomography}
As we mentioned above, in the $\Lambda$CDM model the sub-horizon evolution of gravitational potentials and the matter density fluctuations are described by a single function of time -- the scale-independent growth factor $g(a)$. In models of modified gravity, on the other hand, the dynamics of perturbations can be richer and, generically, the evolution of $\Phi$, $\Psi$ and $\delta$ (the matter density contrast), will be described by different functions of scale and time. Therefore, we expect that different observables will be described by different functions, and by combining different types of measurements, one can try to reconstruct these functions, or at least put a limit on how different they can be. In what follows, we shall give a brief review of the relation between the different types of observables and the gravitational potentials they probe. For a more thorough overview of the various ways of looking for modifications in the growth of perturbations we refer the reader to~\cite{Jain:2007yk}.

\emph{Galaxy Counts} (GC) probe the distribution and growth of matter inhomogeneities. However, to extract the matter power spectrum, one needs to account for the bias, which typically depends on the type of galaxies and can be both time- and scale-dependent. On large scales, where non-linear effects are unimportant, one can use a scale-independent bias factor to relate galaxy counts to the total matter distribution. This relation becomes increasingly complicated and scale-dependent as one considers smaller and smaller scales. In principle, the bias parameters can be determined from higher order correlation functions~\cite{PhysRevLett.73.215,Matarrese:1997sk,Sefusatti:2007ih}. On sub-horizon linear scales, the evolution of matter density contrast is determined by equation~(\ref{density_evol}). Hence, measurements of GC over multiple redshifts can provide an estimate of $\Psi$ as a function of space and time, up to a bias factor. A more direct probe of the potential $\Psi$, would be a measurement of \emph{peculiar velocities}, which follow the gradients of $\Psi$. Such measurements would be independent of uncertainties associated with modeling the bias. 

Peculiar velocity surveys typically use redshift-independent distance indicators to separate the Hubble flow from the local flow, and nearby SNeIa are therefore good candidates; a number of surveys, like the 6dFGS~\cite{Jones:2004zy} and the 2MRS~\cite{Erdogdu:2006nd}, use galaxies. An interesting alternative is offered by the kinetic Sunyaev-Zel'dovich effect in clusters~\cite{Sunyaev:1980nv}, that arises from the inverse Compton scattering of CMB photons off high-energy electrons in the clusters. This effect provides a useful way of measuring the bulk motion of electrons in clusters, hence the peculiar velocity of clusters, but it is limited by low signal-to-noise ratio. Current measurements of peculiar velocities are limited in accuracy, and at this point it is not clear how to forecast the accuracy of future observations. Therefore we did not include them in our observables, even though they are a potentially powerful probe~\cite{Song:2008qt}.

In contrast to galaxy counts and peculiar velocities, which respond to one of the metric potentials,  namely $\Psi$, \emph{Weak Lensing} of distant light sources by intervening structure is determined by spatial gradients of the sum ($\Phi+\Psi$). Hence, measurements of the weak lensing shear distribution over multiple redshift bins can provide an estimate of the space and time variation of the sum of the two potentials. In the $\Lambda$CDM and minimally coupled models of dark energy, the two metric potentials coincide, and therefore WL probes essentially the same growth function that controls the evolution of galaxy clustering and peculiar velocities. In models of modified gravity, however, there could be a difference between the potentials, corresponding to an effective shear component. Measurements of the \emph{Integrated Sachs-Wolfe} effect in the CMB probe the time dependence of the sum of the potentials: $d\l(\Phi+\Psi\r)/dt$. 

By combining multiple redshift information on GC, WL and CMB, and their cross-correlations, one can constrain the differences between the metric potentials and the space-time variation of the effective Newton constant defined in the previous section. Ideally, experimentalists would measure all possible cross-correlations, between all possible pairs of observables, in order to maximize the amount of information available to us. In practice, however, it can be difficult to obtain these cross-correlations, since their measurements require that each of the individual fields (CMB, GC, WL) be measured on the same patch of sky. This will be addressed with near and distant future tomographic large scale structure surveys (such as DES~\cite{DES}, LSST~\cite{LSST} and  PAN-STARRS~\cite{PAN}). Even with conservative assumptions about the data, (i.e. considering only linear scales), it is hoped ~\cite{Zhao:2008bn}, that DES will produce non-trivial constraints on modified growth, and that LSST will do even better (with $\approx 10\%$ relative errors in the parameters).

From the discussions in the previous section, it is clear that the predictions for the growth of structure in models of modified gravity and coupled dark energy can significantly differ from those in $\Lambda$CDM or uncoupled quintessence models. Therefore, LSS and CMB data offer a valuable testing ground for gravity and cosmological tests are expected to play an important role in determining the physics of cosmic acceleration. In principle, a combination of galaxy number counts and weak lensing measurements, along with CMB and SNe data,  will allow a distinction between modified gravity/exotic dark energy models and the standard $\Lambda$CDM model of cosmology. However, as we have already mentioned, it will be harder to break the degeneracy between models of modified gravity and generalized models of dark energy, where the dark fluid is allowed to cluster and carry anisotropic stress. As discussed in~\cite{Jain:2007yk}, the latter  task will require the independent measurements of at least three observables.


\section{Conclusions}
\label{sec:conclusions}
That the expansion rate of the universe is accelerating is now, just over a decade after the first evidence from observations of type Ia supernovae, a firmly established aspect of cosmology. The rapid progress in establishing this fact is a testament to the breathtaking convergence of techniques and technology that has emerged in observational cosmology. In turn, cosmic acceleration has introduced new wrinkles into almost every part of theoretical cosmology, ranging from the details of structure formation, through the CMB and gravitational lensing.

Beyond the fascinating new problems in theoretical cosmology, cosmic acceleration has presented an enormous and, as yet, unmet challenge to fundamental physics. A perfectly good fit to all known cosmological data is that cosmic acceleration is driven by a cosmological constant. However, accounting for the necessary magnitude of such an object seems at least as difficult as attempts to understand why it should be precisely zero. Anthropic arguments, fueled by results from the landscape of string theory and the idea of eternal inflation, may yield a way to understand such a small number. However, it is, at present, far too early to know if this is a sensible outcome of string theory, and there are no developed ideas of how such proposals might be tested. This is not to say, of course, that they are incorrect; merely that there are immense obstacles to establishing such ideas as a tested answer.

If cosmic acceleration is not due to a cosmological constant, then an exotic component of the energy budget of the universe may be the culprit. It is important to recognize up front that any such understanding of cosmic acceleration assumes that the cosmological constant itself is either precisely zero, or at least negligibly small - an unsolved problem in itself. If we make such an assumption, then proposals for a dynamical component of the cosmic energy budget, driving the accelerating expansion, come under the heading of {\it dark energy}. In their simplest forms, ideas of dark energy can be thought of as attempts to construct a model for late-time cosmological inflation, with the key differences being the extremely low energy scales involved, and the fact that this period of acceleration need not end in the way that inflation must. As with early universe inflation, a major challenge is finding a home for the abstract idea within a well-motivated and technically natural particle physics model. In fact, in a number of ways, this problem is more acute for dark energy models, since they must operate in a regime in which we already have a supremely well-tested low-energy effective field theory of much of the matter content of the universe. As we have described, some tentative ideas exist to tackle this problem, but as yet no convincing particle physics implementation of dark energy exists, in contrast to the somewhat natural occurrence of dark matter candidates in theories of physics beyond the standard model.

But dark energy is not the only possible solution to the problem of cosmic acceleration. Instead we might imagine leaving the material contents of our theories untouched, comprised only of the standard model and dark matter, and instead revisiting the dynamical relationship between these components and the evolution of the background geometry of the universe. In other words, we might consider that cosmic acceleration is our first evidence for a modification of General Relativity in the far infrared. One might imagine that this approach would hold out the possibility of a solution to the cosmological constant problem itself, as well as providing an understanding of cosmic acceleration. However, to date this possibility remains unrealized, and just as for dark energy, modified gravity approaches to cosmic acceleration must also merely assume that the cosmological constant is either zero, or negligibly small. 

Making this assumption, a number of authors have considered how one might modify gravity. However, to date all attempts have run into problems, either with matching constraints on gravity at small (solar system) scales, or with the existence of ghost degrees of freedom, raising the possibility of problems similar to those we mentioned when discussing phantom fields. Nevertheless, if one views these models either as a long-wavelength approximation to a better-behaved theory of gravity, or as a very low energy effective theory, in need of an ultraviolet completion to cure its problems, one may extract a variety of predictions for signatures in upcoming cosmological observations.

Perhaps the biggest question to be answered is binary in nature - is cosmic acceleration due to a cosmological constant or not? In many ways, an affirmative answer is the most depressing, since a true cosmological constant varies in neither space nor time, and thus, if it is the driver of acceleration, we already know all that we will ever know about it. As such, we would not expect further observational insights into the required unnaturally small value. More generally, we would like to be able to distinguish among this minimal explanation, a possible dark energy component and the option of an infrared modification of gravity. It has become increasingly clear that our best hope for this is to compare geometrical measures, such as the distance to the surface of last scattering, with measures of the growth of large-scale structure, which depend in detail, in a redshift-dependent manner on the gravitational force law underlying the collapse of overdensities.

In this article we have endeavored to provide an overview of the major approaches to cosmic acceleration and to the techniques through which we hope to further our understanding. What should be clear is that the issues involved here are far from settled, and that the field is hungry for new ideas. It is unclear where such an idea will come from, and for this reason it seems important to pursue all possible avenues, no matter how unpromising they may seem. We hope to have provided a snapshot of some of these avenues of enquiry. It remains to be seen which, if any, of these will provide the key to the problem of cosmic acceleration.

\section*{Acknowledgements}
We would like to thank Rachel Bean, Eanna Flanagan, Justin Khoury and Levon Pogosian for useful discussions during the writing of this article, and all our collaborators for helping develop our understanding. The work of MT is supported in part by National Science Foundation grants PHY-0653563 and PHY-0930521, by Department of Energy grant DE-FG05-95ER40893-A020 and by NASA ATP grant NNX08AH27G. The work of AS is supported by the National Science Foundation under grant AST-0708501.

\section*{References}

\bibliographystyle{unsrt}
\bibliography{rop-bibliography.bib}

\end{document}